\pdfoutput=1

\documentclass[11pt,twoside,a4paper,cmspaper,final,collab]{cms-tdr}
\def\svnVersion{63198:63298}\def\cmsCernNo{2011-065}\def\cmsCernDate{\today}\def\cmsMessage{Submitted to the Journal of High Energy Physics}

\begin{document}\cmsNoteHeader{SUS-10-005}

\hyphenation{had-ron-i-za-tion}
\hyphenation{cal-or-i-me-ter}
\hyphenation{de-vices}
\RCS$Revision: 63222 $
\RCS$HeadURL: svn+ssh://alverson@svn.cern.ch/reps/tdr2/papers/SUS-10-005/trunk/SUS-10-005.tex $
\RCS$Id: SUS-10-005.tex 63222 2011-06-22 14:53:16Z alverson $

\newcommand{\todo}[1]{{\color{magenta}{FIXME: #1}}}
\renewcommand{\fixme}[1]{\todo{#1}}
\newcommand{\sectionauthor}[1]{\texorpdfstring{{\color{blue} #1}}{#1}}
\newcommand{\disclaimer}[1]{\emph{Outlook: #1}}

\newcommand{\update}[1]{{\color{blue}{#1}}}

\newcommand{\T}{\ensuremath{\,\text{T}}\xspace}                % Tesla unit
\newcommand{\meter}{\ensuremath{\,\text{m}}\xspace}            % m unit
\newcommand{\um}{\mu\ensuremath{\,\text{m}}\xspace}            % micron unit
\newcommand{\ns}{\ensuremath{\,\text{ns}}\xspace}              % ns unit
\newcommand{\us}{\mu\ensuremath{\,\text{s}}\xspace}            % ns unit
\newcommand{\mrad}{\ensuremath{\,\text{mrad}}\xspace}          % mrad unit
\newcommand{\urad}{\mu\ensuremath{\,\text{rad}}\xspace}        % micro rad.
\newcommand{\kHz}{\ensuremath{\,\text{kHz}}\xspace}            % kHz unit
\newcommand{\MHz}{\ensuremath{\,\text{MHz}}\xspace}            % MHz unit
\newcommand{\GHz}{\ensuremath{\,\text{GHz}}\xspace}            % GHz unit
\newcommand{\rad}{\ensuremath{\,\text{rad}}\xspace}            %  rad.
\newcommand{\pb}{\ensuremath{\,\text{pb}}\xspace}

\def\antibar#1{\ensuremath{#1\bar{#1}}}
\newcommand{\ppbar}{\antibar{\mathrm{p}}\xspace}
\newcommand{\pp}{\ensuremath{\mathrm{pp}}\xspace}
\newcommand{\e}{\ensuremath{\mathrm{e}}\xspace}
\newcommand{\epm}{\ensuremath{\mathrm{e^{\pm}}}\xspace}
\newcommand{\W}{\ensuremath{\mathrm{W}}\xspace}

\newcommand{\pti}[1]{\ensuremath{p_{\text{T},#1}}\xspace}
\newcommand{\ptgen}{\ensuremath{\pt^{\text{gen}}}\xspace}
\newcommand{\pttrue}{\ensuremath{\pt^{\text{true}}}\xspace}
\newcommand{\ptmin}{\ensuremath{\pt^{\text{min}}}\xspace}
\newcommand{\ptmax}{\ensuremath{\pt^{\text{max}}}\xspace}
\newcommand{\ptreco}{\ensuremath{\pt(\text{recoJet})}\xspace}
\newcommand{\ptave}{\ensuremath{\pt^{\text{ave}}}\xspace}
\newcommand{\ptavemin}{\ensuremath{\pt^{\text{ave,min}}}\xspace}
\newcommand{\ptavemax}{\ensuremath{\pt^{\text{ave,max}}}\xspace}
\newcommand{\ptref}{\ensuremath{\pt^{\text{ref}}}\xspace}
\newcommand{\ppl}{\ensuremath{p_{||}}\xspace}
\newcommand{\ppli}[1]{\ensuremath{p_{||,#1}}\xspace}
\newcommand{\dif}[1]{\ensuremath{\text{d}#1}\xspace}
\newcommand{\mean}[1]{\ensuremath{\langle#1\rangle}}
\newcommand{\tev}{\TeV}
\newcommand{\mht}{\mbox{\ensuremath{\slash\mkern-12mu{H}_{\text{T}}}}\xspace}

\newcommand{\pthat}{\ensuremath{\hat{\text{p}}_\mathrm{T}}\xspace}
\newcommand{\ptrecjet}{\ensuremath{p_{\mathrm{T}}^{\mathrm{recoJet}}}\xspace}
\newcommand{\ptparjet}{\ensuremath{p_{\mathrm{T}}^{\mathrm{particleJet}}}\xspace}
\newcommand{\ptgenjet}{\ensuremath{p_{\mathrm{T}}^{\mathrm{GenJet}}}\xspace}
\newcommand{\ptg}{\ensuremath{p_{\mathrm{T}}^{\gamma}}\xspace}
\newcommand{\ptj}{\ensuremath{p_{\mathrm{T}}^{\mathrm{jet}}}\xspace}
\newcommand{\ptja}{\ensuremath{p_{\mathrm{T}}^{\mathrm{jet1}}}\xspace}
\newcommand{\ptjb}{\ensuremath{p_{\mathrm{T}}^{\mathrm{jet2}}}\xspace}
\newcommand{\ptjc}{\ensuremath{p_{\mathrm{T}}^{\mathrm{jet3}}}\xspace}
\newcommand{\rj}{\ensuremath{R_{\mathrm{jet}}}\xspace}
\newcommand{\rg}{\ensuremath{R_{\mathrm{G}}}\xspace}
\newcommand{\ptiv}[1]{\ensuremath{\mathbf{p}_{\mathrm{T},#1}}\xspace}
\newcommand{\ptivsq}[1]{\ensuremath{|{\mathbf{p}_{\mathrm{T},#1}|}^{2}}\xspace}
\newcommand{\ptivsqm}[1]{\ensuremath{|{\mathbf{p}_{\mathrm{T},#1}^{\mathrm{meas}}|}^{2}}\xspace}
\newcommand{\ptivsqt}[1]{\ensuremath{|{\mathbf{p}_{\mathrm{T},#1}^{\mathrm{true}}|}^{2}}\xspace}
\newcommand{\ptivm}[1]{\ensuremath{\mathbf{p}_{\mathrm{T},#1}^{\mathrm{meas}}}\xspace}
\newcommand{\ptivt}[1]{\ensuremath{\mathbf{p}_{\mathrm{T},#1}^{\mathrm{true}}}\xspace}
\newcommand{\ptjeta}{\ensuremath{\mathbf{p}_{\mathrm{T}}^{\mathrm{jet1}}}\xspace}
\newcommand{\ptjetb}{\ensuremath{\mathbf{p}_{\mathrm{T}}^{\mathrm{jet2}}}\xspace}
\newcommand{\ptjetc}{\ensuremath{\mathbf{p}_{\mathrm{T}}^{\mathrm{jet3}}}\xspace}
\newcommand{\ptjetd}{\ensuremath{\mathbf{p}_{\mathrm{T}}^{\mathrm{jet4}}}\xspace}
\newcommand{\metSM}{\ensuremath{E_{\mathrm{T}}^{\mathrm{Estimated}}}\xspace}
\newcommand{\metC}{\ensuremath{E_{\mathrm{T}}^{\mathrm{C}}}\xspace}
\newcommand{\metbf}{\ensuremath{{\mathbf{E}}_{\mathrm{T}}^{\mathrm{miss}}}\xspace}
\newcommand{\met}{\MET}

\newcommand\defRTDR{\ensuremath{ R2 = \sqrt{[\pi-\Delta\phi(J_{1},\mht) ]^2 +[\Delta\phi(J_{2},\mht) ]^2}    }}
\newcommand\wpj{\ensuremath{\W\textrm{+jets}}\xspace}
\newcommand\zpj{\ensuremath{\Z\textrm{+jets}}\xspace}
\renewcommand\ttbar{\ensuremath{{\rm t\bar{t}}}\xspace}
\newcommand\ttbarMuNu{\ensuremath{\ttbar ( \mu \nu ) {\rm +jets}}\xspace}
\newcommand\wtauhad{\ensuremath{\W \rightarrow \tau_{\rm h} \nu}\xspace}
\newcommand\ttbarTAUHNu{\ensuremath{\ttbar \rightarrow \tau_{\rm h} \nu + {\rm jets}}\xspace}
\newcommand\ttbarDITAUHNu{\ensuremath{\ttbar \rightarrow \tau_{\rm h} \nu + \tau_{\rm h} \nu + {\rm jets}}\xspace}
\newcommand\ttbarMUTAUHNu{\ensuremath{\ttbar ( \mu \nu + \tau_{\mu} \nu) {\rm +jets}}\xspace}
\newcommand\defMHT{\ensuremath{ {\rm MHT} = \mht = | - \sum_{i} \vec{p_{T} (jet_{i}) }|}}
\newcommand{\ztautau}{\ensuremath{\Z \rightarrow \tau \tau}\xspace}
\newcommand{\zmumu}{\ensuremath{\Z \rightarrow \mu^{+} \mu^{-}}\xspace}
\newcommand{\zee}{\ensuremath{\Z \rightarrow \e^{+} \e^{-}}\xspace}
\newcommand{\znunu}{\ensuremath{\Z \rightarrow \nu \bar{\nu}}\xspace}
\newcommand{\zll}{\ensuremath{\Z \rightarrow \ell^{+}\ell^{-}}\xspace}
\newcommand{\wlnu}{\ensuremath{\W \rightarrow \ell\nu}\xspace}
\newcommand{\wenu}{\ensuremath{\W \rightarrow \e\nu}\xspace} 
\newcommand{\wmunu}{\ensuremath{\W \rightarrow \mu \nu}\xspace}
\newcommand{\zellell}{\ensuremath{\Z \rightarrow \ell^{+} \ell^{-}}\xspace}
\newcommand{\znunubr}{\ensuremath{\Z ( \nu \bar{\nu} )}\xspace}
\newcommand{\zellellbr}{\ensuremath{\Z ( \ell^{+} \ell^{-} )}\xspace}
\newcommand{\wmunubr}{\ensuremath{\W ( \mu \nu )}\xspace}
\newcommand{\wenubr}{\ensuremath{\W ( \e\nu )}\xspace}
\newcommand{\wlnubr}{\ensuremath{\W ( \ell \nu )}\xspace}
\newcommand{\wellnubr}{\ensuremath{\W ( \ell \nu )}\xspace}
\newcommand{\wtaunubr} {\ensuremath{\W ( \tau \nu )}\xspace}
\newcommand{\ztautaubr}{\ensuremath{\Z ( \tau \tau)}\xspace}
\newcommand{\zmumubr}  {\ensuremath{\Z ( \mu \mu)}\xspace}
\newcommand{\zeebr}    {\ensuremath{\Z ( \e \e)}\xspace}

\newcommand{\squark}{\sQua}
\newcommand{\gluino}{\sGlu}
\newcommand{\supq}{\sUp}
\newcommand{\sdown}{\sDw}
\newcommand{\sstrange}{\ensuremath{\tilde{s}}\xspace}
\newcommand{\scharm}{\ensuremath{\tilde{c}}\xspace}
\newcommand{\sbottom}{\sBot}
\newcommand{\MEt}{\not\!\! E_{\mathrm{T}}}

\newcommand{\mhtv}[1]{\ensuremath{\slash\mkern-15mu{\vec{H}}_{\text{T}}^{#1}}\xspace}
\newcommand{\METv}{\ensuremath{\slash\mkern-12mu{\vec{E}}_{\text{T}}}\xspace}
\newcommand{\MHT}{\mht}
\newcommand{\MHTv}{\mhtv{}}
\newcommand{\MPT}{\mathrm{MPT}}
\newcommand{\seed}{\mathrm{seed}}
\newcommand{\reco}{\mathrm{reco}}
\newcommand{\gen}{\mathrm{gen}}
\newcommand{\soft}{\mathrm{soft}}
\newcommand{\true}{\mathrm{true}}
\newcommand{\smeared}{\mathrm{smeared}}
\newcommand{\ptcl}{\mathrm{particle}}
\newcommand{\recoJet}{\mathrm{recoJet}}
\newcommand{\genJet}{\mathrm{genJet}}
\newcommand{\ptSoft}{\ensuremath{p_{\mathrm{T,\soft}}}}
\newcommand{\ptSoftV}{\ensuremath{\vec{p}_{\mathrm{T,\soft}}}}
\newcommand{\recoSoft}{\ensuremath{\vec{p}_{\mathrm{T,\soft}}^{\,\reco}}}
\newcommand{\genSoft}{\ensuremath{\vec{p}_{\mathrm{T,\soft}}^{\,\ptcl}}}
\newcommand{\trueSoft}{\ensuremath{\vec{p}_{\mathrm{T,\soft}}^{\,\true}}}
\newcommand{\seedSoft}{\ensuremath{\vec{p}_{\mathrm{T,\soft}}^{\,\seed}}}
\newcommand{\ptV}{\ensuremath{\vec{p}_{\mathrm{T}}}}
\newcommand{\ppp}[3]{\ensuremath{p_{#1,#2}^{#3}}}
\newcommand{\ppt}[2]{\ppp{\mathrm{T}}{#1}{#2}}
\newcommand{\pptV}[2]{\ensuremath{\vec{p}_{\mathrm{T},#1}^{\,#2}}}
\newcommand{\ptInv}{\ptV^{\,\mathrm{invisible}}}
\newcommand{\ptAve}[1]{\langle\pt^{#1}\rangle}
\newcommand{\particleJet}{\mathrm{particleJet}}
\newcommand{\seedJet}{\mathrm{seedJet}}
\newcommand{\seedJets}{\mathrm{seedJets}}
\newcommand{\seedMHT}{\MHT^{\seed}}
\newcommand{\seedMHTv}{\mhtv{\seed}}
\newcommand{\seedHT}{\mathrm{seedHT}}
\newcommand{\partonHT}{\mathrm{partonHT}}
\newcommand{\Dphi}{\Delta\phi}
\newcommand{\DphiMPTMHT}{\Delta\phi(\MPT,\MHT)}
\newcommand{\DphiMHTjet}[1]{\Delta\phi(\MHT,\mathrm{jet\,1\mbox{-}#1})}
\newcommand{\minDphi}{\mathrm{min}\,\Delta\phi(\MHT,\mathrm{jet\,1\mbox{-}3})}
\newcommand{\ssoftT}{\sigma_{\mathrm{T}}^{\soft}}
\newcommand{\ssoftPhi}{\sigma_{\mathrm{\phi}}^{\soft}}
\newcommand{\E}[1]{\times10^{#1}}
\newcommand{\maxL}{\mathrm{max\mbox{-}}\rsL}
\newcommand{\rsN}{(\mathrm{R\&S})}
\newcommand{\effMu}{\varepsilon_{\mu}}
\newcommand{\fakeMu}{\ensuremath{\slash\mkern-12mu{\mu}}\xspace}
\newcommand{\effSplit}{\varepsilon_{\mathrm{split}}}
\newcommand{\ptRel}{\ensuremath{\pt^{\mathrm{rel}}}}
\newcommand{\ptMu}{\ensuremath{\ptV^{\,\mu}}}
\newcommand{\ptJet}{\ensuremath{\ptV^{\,\mathrm{jet}}}}
\newcommand{\footnoteremember}[2]{\footnote{#2}\newcounter{#1}\setcounter{#1}{\value{footnote}}}
\newcommand{\footnoterecall}[1]{\footnotemark[\value{#1}]} 

\newcommand{\minus}{\mbox{-}}
\newcommand{\plus}{\mbox{+}}
\newcommand{\genParticle}{\mathrm{genParticle}}
\newcommand{\minSeedPT}{p_{\mathrm{T,min}}^{\mathrm{seed}}\xspace}
\newcommand{\smearedMHT}{\MHT^{\smeared}}
\newcommand{\smearedMHTv}{\mhtv{\smeared}}
\newcommand{\jet}{\mathrm{jet}}
\newcommand{\rsL}{\mathcal{L}}
\newcommand{\rsLjets}{\rsL_{\mathrm{jets}}}
\newcommand{\rsLsoft}{\rsL_{\mathrm{soft}}}
\newcommand{\rsLcooled}{L_{\mathrm{cooled}}}
\newcommand{\lnLjs}[2]{\minus\ln\rsL_{#1\mathrm{j}+#2\mathrm{s}}^{\max}}
\newcommand{\minLnL}{\min(-\ln\rsL)}
\newcommand{\fcool}{f_{\mathrm{cool}}}
\newcommand{\calo}{\mathrm{calo}}
\newcommand{\caloSoft}{\ensuremath{\vec{p}_{\mathrm{T,\soft}}^{\,\calo}}}
\newcommand{\bisector}{\hat{\eta}}
\newcommand{\along}{\hat{\xi}}
\newcommand{\rs}{\mathrm{R}\mbox{+}\mathrm{S}}
\newcommand{\softRes}{R_{\soft}}
\newcommand{\lfrac}[2]{l_{#1}^{#2}}
\newcommand{\lfracApprox}[2]{l_{#1}^{\prime#2}}
\newcommand{\mufrac}[2]{\mu_{#1}^{#2}}
\newcommand{\mufracApprox}[2]{\mu_{#1}^{\prime#2}}
\newcommand{\br}{\mathrm{BR}}
\newcommand{\brMu}{\br_{\mu}}
\newcommand{\res}{r}
\newcommand{\resLep}{\res_{\mathrm{lep}}}
\newcommand{\resHad}{\res_{\mathrm{had}}}
\newcommand{\resBad}{\res_{\mathrm{bad}}}
\newcommand{\resGood}{\res_{\mathrm{good}}}
\newcommand{\resLepApprox}{\resLep^{\prime}}
\newcommand{\resHadApprox}{\resHad^{\prime}}
\newcommand{\distLep}{h_{\mathrm{lep}}^{\prime}}
\newcommand{\distHad}{h_{\mathrm{had}}^{\prime}}
\newcommand{\gausGood}{\mathcal{G}_{\mathrm{good}}}
\newcommand{\fracBad}{w_{\mathrm{bad}}}
\newcommand{\fracGood}{w_{\mathrm{good}}}
\newcommand{\dope}[2]{d_{#1}^{#2}}
\newcommand{\dopeApprox}[2]{d_{#1}^{\prime#2}}
\newcommand{\dopeMu}{d_{n,\mu\mbox{-}\mathrm{tag}}}
\newcommand{\dijet}{\mathrm{dijet}}
\newcommand{\ndof}{\mathrm{ndof}}
\newcommand{\METvJ}{\ensuremath{\slash\mkern-12mu{\,\vec{J}}_{\text{T}}}\xspace}
\newcommand{\badFrac}{\ensuremath{f^{\mathrm{ECAL}}_{\mathrm{masked}}}\xspace}

\newcommand{\DeltaPhi}{\ensuremath{\Delta\phi_{\text{min}}}}

\renewcommand{\GeVc}{\GeV}
\renewcommand{\GeVcc}{\GeV}

\newcommand{\RplusS}{R\&S\xspace}

\cmsNoteHeader{SUS-10-005} % This is over-written in the CMS environment: useful as preprint no. for export versions
\title{Search for New Physics with Jets and Missing Transverse Momentum in $\pp$ Collisions at $\sqrt{s} = 7 \TeV$}% Force line breaks with \\

\author{The CMS Collaboration}

\date{\today}

\abstract{
A search for new physics is presented based on an event signature of at least three jets accompanied by large missing transverse momentum, using a data sample corresponding to an integrated luminosity of $36\pb^{-1}$ collected in proton--proton collisions at $\sqrt{s}=7\TeV$ with the CMS detector at the LHC.
No excess of events is observed above the expected standard model backgrounds, which are all estimated from the data. Exclusion limits are presented for the constrained minimal supersymmetric extension of the standard model. Cross section limits are also presented using simplified models with new particles decaying to an undetected particle and one or two jets.
}

\hypersetup{%
pdfauthor={CMS Collaboration},%
pdftitle={Search for New Physics with Jets and Missing Transverse Momentum in pp collisions at sqrt(s) = 7 TeV},%
pdfsubject={CMS},%
pdfkeywords={CMS, physics, SUSY, jets, MET}}

\maketitle %maketitle comes after all the front information has been supplied

\section{Introduction}
\label{sec:introduction}

Several theories beyond the standard model (SM) of particle physics address the gauge hierarchy problem and other shortcomings of the SM by introducing a spectrum of new particles that are partners of the SM particles~\cite{SUSY0,LittleHiggs,UED}. These new particles may include neutral, stable, and weakly interacting particles that are good dark-matter candidates. The identity and properties of the fundamental particle(s) that make up dark matter are two of the most important unsolved problems in particle physics and cosmology. The energy density of dark matter is approximately five times larger than for the normal baryonic matter that corresponds to the luminous portion of the universe.
A review on dark matter can be found in Ref.~\cite{Feng:2010gw}.

Many dark-matter candidates are stable as a result of a conserved quantity. In supersymmetry (SUSY) this quantity is $R$ parity, and its conservation requires all SUSY particles to be produced in pairs and the lightest SUSY particle (LSP) to be stable. Coloured SUSY particles can be pair-produced copiously at the Large Hadron Collider (LHC). These particles will decay directly into SM particles and an LSP or via intermediate colour-singlet states that ultimately decay into an LSP, resulting in a large amount of energy deposited in the detector.  The LSP will pass through the detector without interacting, carrying away a substantial amount of energy and creating an imbalance in the measured transverse momentum (\pt).

Experiments at the Tevatron~\cite{CDFLimits,D0Limits,Abazov200934}, SPS~\cite{UA1Limits,UA2Limits}, LEP~\cite{ALEPHSUSY,DELPHISUSY,L3SUSY,OPALSUSY}, and HERA colliders~\cite{:2006je,Aid:1996es} have performed extensive searches 
for SUSY and set lower limits on the masses of SUSY particles. At the LHC, the CMS Collaboration has previously published limits in the all-hadronic channel based on a search using the $\alpha_{\rm T}$~\cite{Randall:2008rw}
kinematic variable~\cite{RA1}. The ATLAS Collaboration has also published limits from a missing transverse momentum and multijet search~\cite{ATLASJetMET}.

In this paper, results are presented from a search for large missing transverse momentum in multijet events produced in {\rm pp} collisions at a centre-of-mass-energy of $7 \TeV$, using a data sample collected with the CMS detector at the LHC in 2010, corresponding to an integrated luminosity of $36 \pbinv$. The results of the search are presented in the context of the constrained minimal supersymmetric extension of the standard model (CMSSM)~\cite{CMSSM}, and in the more general context of simplified models~\cite{Alwall:2008ag,Alwall:2008zz,SMS}. These latter models are designed to characterize experimental data in terms of a small number of particles whose masses and decay branching fractions are allowed to vary freely.
The results are independent of any more complete
theory that addresses the deeper problems of particle physics, yet they can be translated into any such desired framework.

This search is complementary to the CMS analysis~\cite{RA1} that used the kinematic variable $\alpha_{\rm T}$ as the search variable in events with at least two jets. That variable is very effective in suppressing the QCD multijet background but with some loss of signal acceptance. In contrast, this search only selects events with $\ge 3$ jets, and the missing and visible transverse momentum sums are used as search variables for an inclusive selection with a higher signal acceptance.

The main backgrounds in this analysis are: (a) an irreducible background from \zpj events, with the \Z boson decaying to $\nu\bar{\nu}$, denoted as $\znunubr$+jets; (b) $\W$+jets and $\ttbar$ events, with either the directly-produced $\W$ boson or one of the $\W$ bosons from the top-quark decays going directly or via a $\tau$ to an $\e$ or $\mu$ that is lost, or going to a $\tau$ that decays hadronically. In all these cases, one or more neutrinos provide a genuine source of missing transverse momentum; and (c) QCD multijet events with large missing transverse momentum from leptonic decays of heavy-flavour hadrons inside the jets, jet energy mismeasurement, or instrumental noise and non-functioning detector components. The relative contributions of these three categories of backgrounds depend on the event selection.

This paper is organized as follows. The CMS detector and event reconstruction are described in Section~\ref{sec:detector}. In Section~\ref{sec:eventselection}, the event selection criteria are presented. The backgrounds to this search are directly determined from the data, in some cases with novel techniques which are being applied here for the first time. In Section~\ref{sec:zinv}, the irreducible $\znunubr$+jets background is estimated from $\gamma$+jets events, and alternative \Z and \W control samples are studied. The background from $\W$+jets and $\ttbar$ where a lepton is either lost or is a hadronically decaying tau lepton is estimated from $\mu$+jets events by ignoring or replacing the muon, as discussed in Section~\ref{sec:wtop}. The QCD multijet kinematics are predicted using measured jet resolution functions to smear events obtained by a procedure that produces well-balanced events out of inclusive multijet data, as discussed in Section~\ref{sec:qcd}. As a cross-check, the correlation between the transverse missing momentum vector and the angular distance between that vector and the closest leading jet is used to predict the tail of the missing-momentum distribution. In Section~\ref{sec:sensitivity}, the interpretation of the observed data is presented.

\section{The CMS detector and event reconstruction}
\label{sec:detector}

The central feature of the CMS apparatus is a superconducting solenoid $13 \meter$ in length and 
$6 \meter$ in diameter, which provides an axial magnetic field of $3.8 \T$. The bore of the solenoid 
is instrumented with various particle detection systems.  The steel return yoke outside the 
solenoid is in turn instrumented with gas detectors which are used to identify muons.  
Charged particle trajectories are measured by the silicon pixel and strip tracker, covering 
$0 < \phi < 2\pi$ in azimuth and $|\eta| < 2.5$, where the pseudorapidity $\eta$ is defined 
as $\eta = -\ln \left[ \tan (\theta/2) \right]$, with $\theta$ being the polar angle of the particle's momentum with respect to the counterclockwise beam direction. 
A lead-tungstate crystal electromagnetic calorimeter (ECAL) and a brass/scintillator hadronic 
calorimeter (HCAL) surround the tracking volume and cover the region $|\eta| < 3$. Quartz/steel %\v{C}erenkov-radiation-based
forward hadron calorimeters extend the coverage to $|\eta|\le 5$.
The detector is nearly hermetic, allowing for momentum balance measurements in the plane 
transverse to the beam directions. A detailed description of the CMS detector can be found 
elsewhere \cite{CMS}.

All physics objects are reconstructed with a particle-flow technique~\cite{PFT-09-001}. 
This algorithm identifies and reconstructs individually the particles produced in the collision, namely charged and neutral hadrons, photons, muons, and electrons, by combining the information from the tracking system, the calorimeters, and the muon system. 
All these particles are clustered into jets using the anti-$k_T$ algorithm 
with a distance parameter of $0.5$~\cite{antikt} from {\sc FastJet}~\cite{Cacciari:2006sm}. Jet energies are corrected for the non-linear calorimeter response using calibration factors derived from simulation, 
and, for jets in data, an additional residual energy correction derived from data is applied~\cite{PAS-JME-10-010}. As the average number of additional pileup interactions during the LHC 2010 data taking is roughly between two and three, no subtraction of the pileup energy deposits is performed.

\section{Sample selection}
\label{sec:eventselection}

The event selection for this search aims to be inclusive, such that it can detect new physics from any model yielding a high-multiplicity hadronic final state
  with missing transverse momentum. Therefore, the observables of central interest in the
  search are chosen to be the magnitude of the missing transverse momentum \MHT calculated from jets, and the scalar sum of the jet transverse momenta \HT. The choice of these observables and the applied background suppression cuts
aim for a minimal kinematic bias in the search for new physics signals. This facilitates the characterization of new physics in the case of a discovery.
Furthermore, the selection is chosen to be efficient for models containing new particles with sufficiently small mass and thus sizeable production yield for the integrated luminosity used in this search.
In this section, the event selection is described, based on the above considerations.

\subsection{Trigger selection and cleaning of the data sample} \label{sec:evseldata}

The data used in this analysis were collected with triggers based on the quantity $\HT^{\rm trig}$, defined as the scalar sum of the
transverse momenta of reconstructed calorimeter jets (without response corrections) having $\pt > 20 \GeVc$ and $|\eta| < 5$. %~\footnote{At the start of the run, for a limited time equivalent to an integrated luminosity of $0.18\pbinv$, the minimal jet \pt threshold was at $30 \GeVc$.},
The $\HT^{\rm trig}$ threshold varied between $100$ and $150 \GeV$ as the instantaneous luminosity of the LHC increased. The
\HT trigger has a high acceptance for low-mass hadronic, new-physics signatures,
and it enables the simultaneous collection of several control samples used to estimate the backgrounds. The trigger
efficiency as a function of the particle-flow-based \HT, defined below in Section~\ref{sec:evselcuts}, is found to be close to 100\%
for \HT values above $300$ \GeV.

Ways to remove events with a poor \MHT measurement were investigated using both simulation and data.
Various sources of noise in the electromagnetic and hadronic calorimeters are rejected~\cite{ecalnoise,METJINST}.
Beam-related background events and displaced satellite collisions are removed by requiring a well-reconstructed primary vertex within the luminous region, applying a beam-halo veto~\cite{METJINST}, asking for a significant fraction of tracks in the event to be of high quality, and requiring the scalar sum of the transverse momenta of tracks associated with the primary vertex to be greater than 10\% of the scalar sum of the transverse momenta of all jets within the tracker acceptance.
Events are also rejected in which a significant amount of energy is determined to have been lost in the approximately 1\% of non-functional crystals in the ECAL that are masked in reconstruction~\cite{ecalnoise}.  Such losses are identified either by exploiting the
energy measured through a parallel readout path used for the online trigger, or by measuring the energy deposited around masked crystals when information from this parallel readout path is not available. The small inefficiency for signal events induced by this cleaning is discussed further in Section~\ref{sec:resultslimits}.

\subsection{Baseline and search event selections} \label{sec:evselcuts}

The search selection starts from a loosely selected sample of candidate events. From this so-called baseline sample, tighter search selection criteria are then applied to obtain the final event sample. The baseline selection requirements are:

\begin{itemize}
\item At least three jets with $\pt> 50\GeVc$ and $| \eta | < 2.5$.
\item $\HT > 300 \GeV$, with \HT defined as the scalar sum of the transverse momenta of all jets with $\pt>50 \GeVc$ and $|\eta|<2.5$.
\item $\mht> 150\GeV$, with \mht defined as the magnitude of the negative vector sum of the transverse momenta of all jets with $\pt>30 \GeVc$ and $| \eta | < 5$. This requirement suppresses the vast majority of the QCD multijet events.
\item $ | \Delta\phi ( J_{n} , \mht ) | > 0.5$, $n=1,2$ and $ | \Delta\phi ( J_{3} , \mht ) | > 0.3$, where $\Delta\phi$ is the azimuthal angular difference between the jet axis $J_{n}$ and the \mht direction for the three highest-\pt jets in the event.
This requirement rejects most of the QCD multijet events in which a single mismeasured jet yields a high-\mht value.
\item No isolated muons or electrons in the event. A loose lepton definition is employed to reject the leptonic final states of \ttbar and $\W/\Z$+jets events.
Muons and electrons are required to have $\pt > 10 \GeVc$ and produce a good quality track that is matched to the primary vertex within $200\mum$ transversely and $1\cm$ longitudinally. They must also be isolated, requiring a relative isolation variable to satisfy:
{\setlength\abovedisplayskip{4pt plus 3pt minus 7pt} \setlength\belowdisplayskip{3pt plus 3pt minus 7pt} \[ \mbox{\small $\left[ \sum^{\Delta R<0.4}\pt{}^{\mathrm{charged~hadron}} +\sum^{\Delta R<0.4}\pt{}^{\mathrm{neutral~hadron}} +\sum^{\Delta R<0.4}\pt{}^{\rm photon} \right] / \pt^{\rm lepton}$ } < 0.2 , \]  }
\hspace{-1mm}where $\pt{}^{\mathrm{charged~hadron}}$, $\pt{}^{\mathrm{neutral~hadron}}$, and $\pt{}^{\rm photon}$ are, respectively, the momentum of charged hadrons, neutral hadrons, and photons in the event within a distance $\Delta R = 0.4$ in $\eta$--$\phi$ space of the lepton.
Muons are required to have $|\eta|<2.4$, whereas electrons must have $|\eta| < 2.5$, excluding the barrel-endcap transition region $1.44 < |\eta| < 1.57$.
\end{itemize}

Two search regions are chosen, based on the observables central to this inclusive jets-plus-missing-transverse-momentum search. The first selection, defining the high-\MHT search region, tightens the baseline cuts with an $\MHT > 250\GeV$ requirement, motivated by the search for a generic dark-matter candidate, which gives a large background rejection. The second selection adds a cut of $\HT > 500\GeV$ to the baseline criteria, yielding the high-\HT search region, which is sensitive to the higher multiplicities from cascade decays of high-mass new-physics particles. Such cascades lead to more energy being transferred to visible particles and less to invisible ones.

\subsection{Data--simulation comparison} \label{sec:evselmc}

Several Monte Carlo (MC) simulation samples are used, produced with a detailed CMS detector simulation based on {\sc Geant4}~\cite{Geant2}.
Samples of QCD multijet, \ttbar, \W/\zpj, $\gamma$+jets, diboson, and single-top events were generated with the \PYTHIA{}6~\cite{pythia} and \MADGRAPH~\cite{madgraph} generators using the CTEQ6.1L~\cite{Nadolsky:2008zw} parton distribution functions.
For the \ttbar background an approximate next-to-next-to-leading-order
(NNLO) cross section of $165\pb$~\cite{Kidonakis:2010dk} is used, while
the cross sections for \wlnubr{}+jets ($31\,300\pb$) and \znunubr{}+jets ($5\,769\pb$) are
derived from an NNLO calculation with FEWZ~\cite{Melnikov:2006kv}.
While already excluded~\cite{RA1}, the LM1 CMSSM point~\cite{PhysTDR2} is used as a benchmark for new physics in this search. This point has a cross section of $6.5\pb$ at NLO, calculated with {\sc Prospino}~\cite{Prospino}. It is defined to have a universal scalar mass $m_0 = 60 \GeVcc$, universal gaugino mass $m_{1/2} = 250 \GeVcc$, universal trilinear soft SUSY-breaking-parameter $A_0 = 0$, the
ratio of the vacuum expectation values of the two Higgs doublets $\tan \beta = 10$, and the sign of the Higgs mixing parameter ${\rm sign}(\mu)$ positive. The squark and gluino masses for LM1 are respectively $559 \GeVcc$ and $611 \GeVcc$, and the LSP mass is $96\GeVcc$.

\begin{figure}[htb!]
  \begin{center}
\hspace*{-5mm}
\includegraphics[width=0.52\textwidth,height=0.52\textwidth]{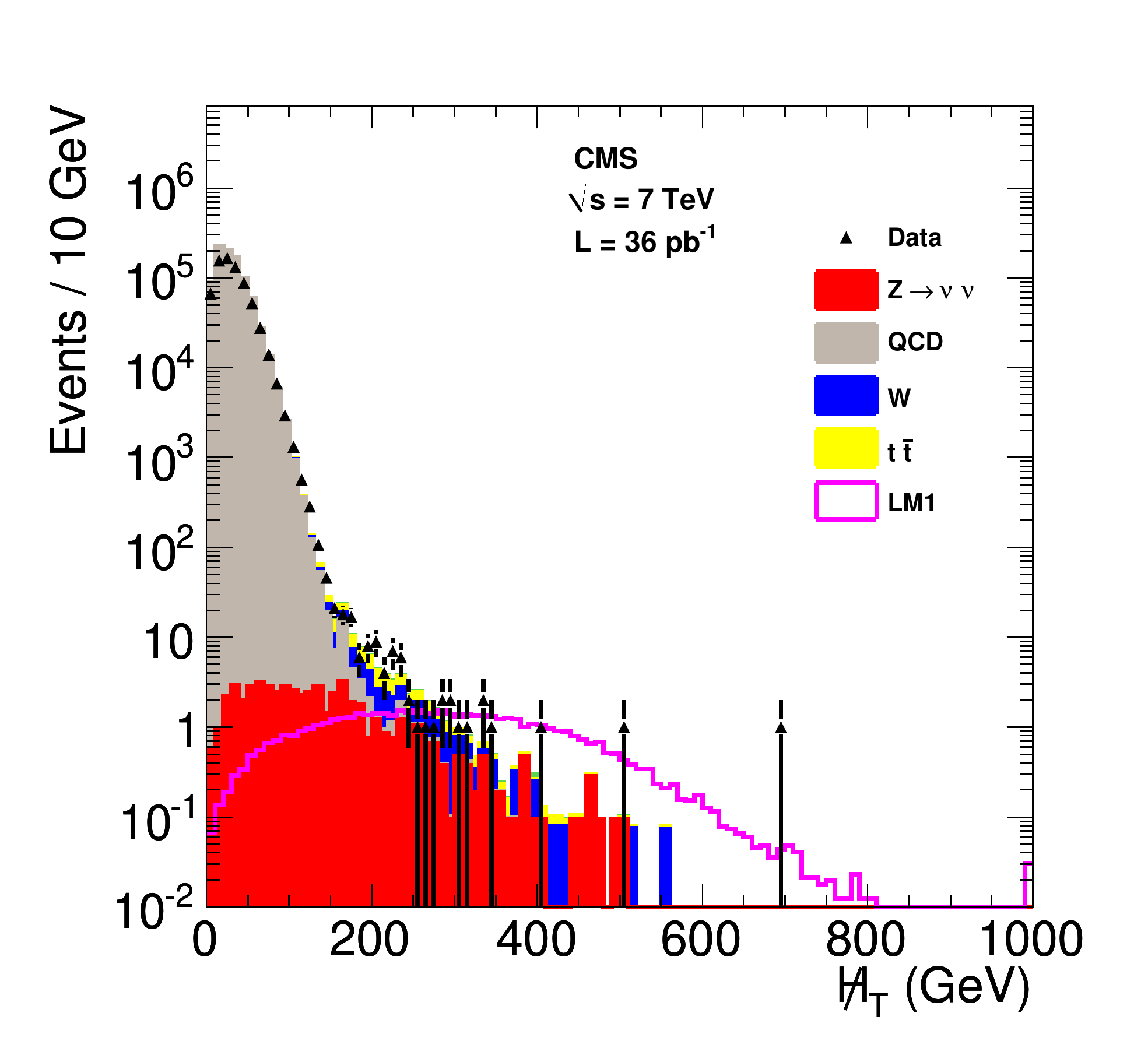}
\hspace*{-5mm}
\includegraphics[width=0.52\textwidth,height=0.52\textwidth]{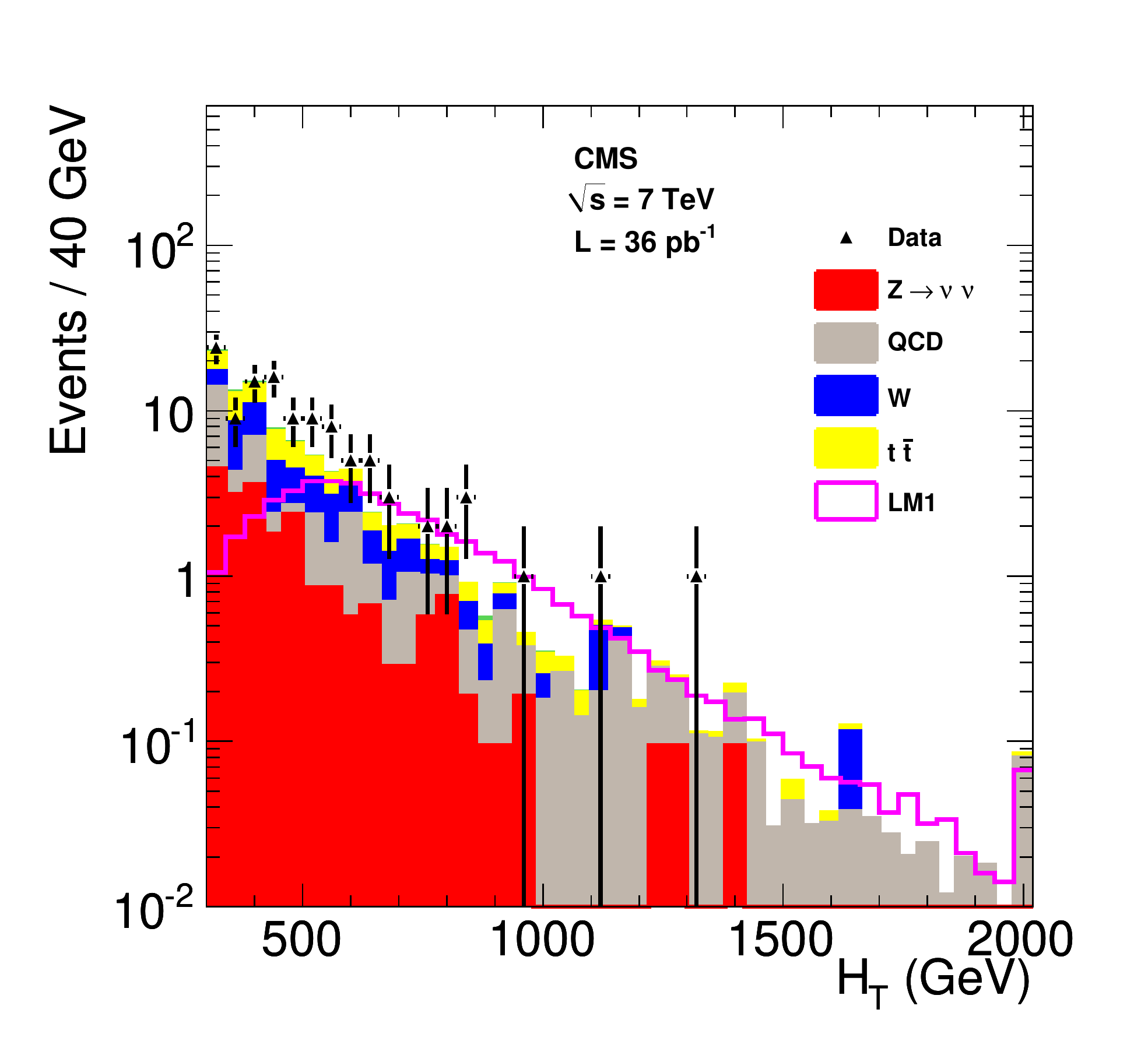}
\hspace*{-7mm}
    \caption{The (left) \MHT and (right) \HT distributions for the data and MC simulation samples with all baseline selection cuts applied except the \MHT and \HT requirements, respectively. The distributions for the individual backgrounds are shown separately, along with the predicted distributions for the LM1 SUSY point. However, these simulated distributions are not used to estimate the backgrounds in this analysis. Instead, the backgrounds are determined directly from the data.}
    \label{fig:evselMHTandHT}
  \end{center}
\end{figure}

The event yields in the data and the simulated samples after two loose selections, the baseline selection, and the two different search event selections are summarized in Table~\ref{tab:DataSignalBkg}, where the simulated event yields correspond to an integrated luminosity of $36 \pbinv$.
The \MHT and \HT distributions for data and MC simulation are compared in Fig.~\ref{fig:evselMHTandHT} after the baseline selection.
In the following sections, however, all the backgrounds in this search are estimated directly from data.

\begin{table}[htb]
  \centering
  \caption{Event yields in data and simulated samples were produced for five different selection criteria. The latter are normalized to an integrated luminosity of $36 \pbinv$. All simulated samples were generated with the \PYTHIA and \MADGRAPH generators. The row labeled LM1 gives the expected yield for the benchmark supersymmetric model described in the text.}
  \label{tab:DataSignalBkg}
  \begin{tabular}{|l|ccccc|}
      \hline \rule{0pt}{12pt}%
        & Baseline & Baseline & Baseline & High-\MHT & High-\HT \\
        & no $\Delta\phi$ cuts & no $\e/\mu$ veto & selection & selection & selection \\
        & no $\e/\mu$ veto & & & & \\
       \hline \rule{0pt}{12pt}%
       LM1                                               &  71.2 &  60.4 &  45.0 & 31.3 & 33.8 \\
       \hline \rule{0pt}{12pt}%
       QCD multijet                                      & 222.0 &  27.0 &  24.6 &  0.2 &  9.9 \\
       $\znunubr$+jets                                   &  26.7 &  21.1 &  21.1 &  6.3 &  5.7 \\
       $\wlnubr$+jets                                    &  93.9 &  57.8 &  23.5 &  4.7 &  7.6 \\
       $\ttbar$                                          &  57.5 &  40.1 &  21.9 &  2.6 &  5.7 \\
       \parbox{4cm}{\W{}\W{}+\W{}\Z{}+\Z{}\Z{}+${\rm t}$\W \\
       +\W{}$\gamma$+\Z{}$\gamma$+$\Z{}/\gamma^{\star}$} &   6.1 &   3.4 &   2.1 &  0.2 &  0.2 \\
       \hline \rule{0pt}{12pt}%
       Total MC background                               & 406   & 149   &  93   & 14   & 29   \\
       Data                                              & 482   & 180   & 111   & 15   & 40   \\
       \hline
  \end{tabular}
\end{table}

\section{\texorpdfstring{\znunubr{}+jets}{Z(vv)+jets} background estimation}
\label{sec:zinv}

The production of a \Z boson and jets, followed by the decay of the \Z boson into neutrinos, constitutes an irreducible background.
The first method to estimate this background from the data exploits the electroweak correspondence between the \Z boson and the photon at high \pt, where they exhibit similar characteristics, apart from electroweak coupling differences and asymptotically vanishing residual mass effects~\cite{PAS-SUS-08-002}. The cross-section ratio between the \Z-boson and photon production provides a robust prediction of the missing transverse momentum spectrum for invisible \Z{} bosons at high~\pt, where the photon production cross section is asymptotically about 20\% less than the one for inclusive \Z-boson production.
One important distinction between photon and \Z-boson production arises from the breakdown of the leading-order calculation of the $\gamma$+jets process for small-angle or vanishing-energy emission of the photon in the absence of a mass to regularize the resulting divergences. This can be mitigated by imposing isolation requirements on the selected photon sample.

The $\gamma$+jets control sample is collected using single-photon triggers, which were measured to be fully efficient for events passing the baseline selection. In the offline selection, photon candidates are distinguished from electrons by a veto on the presence of a track seed in the pixel detector. Photons from QCD multijet events are suppressed by requiring them to be isolated and the shower shape in the $\eta$ coordinate to be consistent with that of a single photon~\cite{CMS-PAS-EGM-10-006}.

For the derivation of the \Z/$\gamma$ cross-section correction factor, simulated $\gamma$+jets and \znunu \MADGRAPH samples are used, in addition to the QCD multijet, \W/\zpj, and \ttbar samples.
The contribution of fragmentation photons, which do not have a counterpart in the massive \Z-boson production, is estimated from NLO {\sc JetPhox}~\cite{jetphox} calculations to be $(5 \pm 1)\%$~\cite{Khachatryan:2010fm} in the selected photon sample. A second background arises from isolated neutral pions and $\eta$ mesons decaying to pairs of secondary photons. For high-momentum mesons, these photon pairs are sufficiently well collimated to be reconstructed as a single photon. Using a method that fits a photon isolation observable to the expected distributions for real and background photons, %~\cite{CMS-PAS-QCD-10-037}, cannot quote unfinished PAS/paper
the purity of the prompt photon sample is found to be $(94^{+6}_{-9})\%$ after the baseline selection, which is in good agreement with simulation. Finally, the background from electrons mis-identified as photons is measured with \zee data events and is found to be negligible for the search selections.

In order to predict the number of \znunubr{}+jets events passing the search selections, the selected $\gamma$+jets control sample needs the following corrections after the background subtraction. 
First, the cross-section ratio between the \znunubr{}+jets and $\gamma$+jets processes is estimated from simulation. The photon selection and isolation cuts are applied to the simulated samples when estimating this correction factor, hence folding the detector acceptance correction into this \Z/$\gamma$ correspondence. The correction factors for the baseline, the high-\MHT, and the high-\HT selections are $0.41\pm0.03$, $0.48\pm0.06$, and $0.44\pm0.06$, respectively, where the uncertainties are statistical only. The uncertainty on the acceptance is taken as 5\%~\cite{RA1}, while the theoretical uncertainty is estimated from a comparison of leading to next-to-leading-order calculations of the ratio of \Z{} and $\gamma$ production with two jets~\cite{Bern:2011pa}.
This dedicated calculation was performed for the different selections in this analysis adapted to only two jets. The addition of an extra jet is not expected to induce a significant effect. This leads to a 10\% theoretical uncertainty on the $\Z$/$\gamma$ cross-section ratio for the baseline selection, which is taken as a uniformly distributed systematic uncertainty with a standard deviation of 6\%. 
The photon reconstruction inefficiency is estimated in Ref.~\cite{Khachatryan:2010fm} to be $(3.5 \pm 1.4)\%$.
Finally, the photon identification and isolation efficiency is corrected for the difference between data and simulation. The correction is determined~\cite{CMS-PAS-EGM-10-006} to be $1.01\pm 0.02$, after baseline selection.

In Table~\ref{tab:photoncorrsyst} the full list of corrections is summarized for the baseline and search selections, along with the corresponding systematic uncertainties. The results for the \znunubr{}+jets prediction from the $\gamma$+jets control sample are summarized in Table~\ref{tab:photonresults}. The prediction is in good agreement with the one found directly from the MC simulation, also given in in Table~\ref{tab:photonresults}.

\begin{table}[htb]
\centering
\caption{Overview of all correction factors and corresponding systematic uncertainties for the prediction of the \znunubr+jets background from the $\gamma$+jets control sample for each of the selections.}
\label{tab:photoncorrsyst}
\begin{tabular}{|r@{\hspace{3mm}$\pm$}l|r@{\hspace{3mm}$\pm$}lr@{\hspace{3mm}$\pm$}lr@{\hspace{3mm}$\pm$}l|}
\hline
\multicolumn{2}{|c|}{\rule{0pt}{12pt}} & \multicolumn{2}{c}{Baseline} & \multicolumn{2}{c}{High-\MHT} & \multicolumn{2}{c|}{High-\HT} \\
\multicolumn{2}{|c|}{}                 & \multicolumn{2}{c}{selection} & \multicolumn{2}{c}{selection} & \multicolumn{2}{c|}{selection} \\
\hline \rule{0pt}{12pt}%
\Z/$\gamma$ correction & theory          & $0.41$ & 6\,\% & $0.48$ & 6\,\% & $0.44$ & 4\,\% \\
                       & acceptance &    &  5\,\% &       &  5\,\% &       &   5\,\% \\
                       & MC stat.   &    &  7\,\% &       & 13\,\% &       &  13\,\% \\
\multicolumn{2}{|l|}{Fragmentation}      & $0.95$ & $1$\,\%
                                         & $0.95$ & $1$\,\%
                                         & $0.95$ & $1$\,\% \\
\multicolumn{2}{|l|}{Secondary photons}  & $0.94$ & $9$\,\%
                                         & $0.97$ & $10$\,\%
                                         & $0.90$ & $9$\,\% \\
\multicolumn{2}{|l|}{Photon mistag}      & $1.00$ & $1$\,\%
                                         & $1.00$ & $1$\,\%
                                         & $1.00$ & $1$\,\% \\
\multicolumn{2}{|l|}{\parbox{5cm}{Photon identification and\\isolation efficiency}}
                                         & $1.01$ & $2$\,\%
                                         & $1.01$ & $2$\,\% 
                                         & $1.01$ & $2$\,\% \\
\hline
\multicolumn{2}{|l|}{\rule{0pt}{12pt}Total correction}
                                         & $0.37$ & $14$\,\%
                                         & $0.45$ & $18$\,\%
                                         & $0.38$ & $17$\,\% \\
\hline
\end{tabular}
\end{table}

\begin{table}[htb]
\centering
\caption{Number of $\gamma$+jets events in the data and the resulting estimate of the \znunubr{}+jets background, as well as the prediction from the MC simulation, for each of the selections, with their statistical and systematic uncertainties. The estimate from data is obtained by multiplying the number of events in the $\gamma$+jets sample with the total correction factor from Table~\ref{tab:photoncorrsyst}.}
\label{tab:photonresults}
\begin{tabular}{|l|l|l|l|}
\hline \rule{0pt}{12pt}%
 & \multicolumn{1}{c|}{Baseline} & \multicolumn{1}{c|}{High-\MHT} & \multicolumn{1}{c|}{High-\HT} \\
 & \multicolumn{1}{c|}{selection} & \multicolumn{1}{c|}{selection} & \multicolumn{1}{c|}{selection} \\
\hline \rule{0pt}{12pt}%
 $\gamma$+jets data sample & $72  $ & $16$ & $22$ \\
 \znunu estimate from data & $26.3 \pm 3.2 \pm 3.6 $
                           & \;\;$ 7.1 \pm 1.8 \pm 1.3 $ 
                           & \;\;$ 8.4 \pm 1.8 \pm 1.4 $\\
 \znunu MC expectation     & $21.1 \pm 1.4$ & $ \;\;6.3 \pm 0.8$ & $ \;\;5.7 \pm 0.7$\\
\hline
\end{tabular}
\end{table}

A potential alternative method to estimate the \znunubr{}+jets background in a conceptually more straightforward way uses \zellellbr{}+jets data events. 
By counting the pair of leptons as missing transverse momentum, the topology of the \znunu process can be reproduced, and all jet-related selection criteria can be directly applied.
Only a small number of \zellellbr{}+jets events pass the selection criteria in the currently available data.
After the baseline selection, applying $\zellell$ selection requirements and  correcting for the acceptance, efficiencies, and different branching fractions, 
the predicted \znunu rates are found to be compatible with the simulation predictions within uncertainties.
However, none of the \zee and \zmumu events pass either of the search selections.

More events can be used for predicting the $\znunubr$+jets background by using $\wellnubr$+jets events. This third method requires additional corrections for the $\W$-$\Z$ correspondence and the \ttbar contamination in the $\ell$+jets control sample. With the available data, a few events are selected in the control samples for the search regions. The predicted number of $\znunubr$+jets background events from this method is consistent with the predictions from the $\gamma$+jets events and the simulation.

\section{\texorpdfstring{\W}{W} and \texorpdfstring{\ttbar}{top} background estimation}
\label{sec:wtop}

The muon and electron vetoes described in Section~\ref{sec:evselcuts} aim to suppress SM events with an isolated lepton. The $\wpj$ and $\ttbar$ events, however, are not rejected by this lepton veto when a
lepton from a \W or top-quark decay is outside the geometric or kinematic acceptances, not reconstructed, not isolated (these three cases are denoted as a ``lost lepton''), or is a tau lepton that decays hadronically (denoted as $\tau_{\rm h}$).
In this section, two
methods are presented to estimate these two components
of the $\wpj$ and $\ttbar$ backgrounds from data.
The first method uses a $\mu$+jets control sample, after correcting for lepton inefficiencies, to estimate the
number of events that fail the isolated lepton reconstruction. The
other method predicts the hadronic $\tau$ background from a similar $\mu$+jets control
sample by substituting a $\tau$ jet for the muon.
For both methods the chosen $\mu$+jets control sample fully represents the
hadronic and other properties of the background it predicts.

The sum of the lost-lepton and hadronic-$\tau$ predictions yields an estimate for the sum of the $\wpj$ and $\ttbar$ background.
The \ttbar contribution is also measured separately as a cross-check. The method predicts the \ttbar background 
from a ${\rm b}$-tagged control sample by correcting for the ${\rm
b}$-tag efficiency, acceptance, and the residual \Z, \W, and multijet contamination.
Using the \W{}-to-\ttbar ratio predicted by simulation, the result is found to be consistent with the estimates described in the subsequent sections.

\subsection{The \texorpdfstring{$\W/\ttbar\to\e,\mu\textrm{+X}$}{W/tt->e,mu+X} background estimation} 
\label{sec:wtop_lostlepton}
 
The background from \W{}+jets and \ttbar events, where a \W boson decays into a muon or an electron that is not rejected by the explicit lepton veto, is measured using a muon control sample.
This control sample is selected by requiring exactly one muon that is isolated and passes the identification quality cuts discussed in Section~\ref{sec:evselcuts}.
From simulation, more than 97\% of this sample are \W{}+jets and \ttbar events.
In order to estimate the number of events in the signal region with non-isolated, but identified electrons and muons,
events in the isolated-muon control sample are weighted according to
$\left( \frac{\epsilon_{\rm ID}^{\e,\mu}}{\epsilon_{\rm ID}^{\mu}} \right) \left( \frac{1-\epsilon_{\rm  ISO}^{\e,\mu}}{\epsilon_{\rm  ISO}^\mu} \right)$, where $\epsilon_{\rm ISO}^{\e,\mu}$ are the electron and muon isolation efficiencies and $\epsilon_{\rm  ID}^{\e,\mu}$ the corresponding identification efficiencies.
To model the number of events in the signal sample containing non-identified electrons or muons, the control sample is corrected by the factor $\left( \frac{1}{\epsilon_{\rm ISO}^{\mu}} \right)
\left( \frac{1-\epsilon_{\rm ID}^{\e,\mu}}{\epsilon_{\rm ID}^{\mu}} \right)$.

The lepton isolation efficiency is measured from $\zellell$ events using a tag-and-probe method~\cite{Khachatryan:2010xn} as a function of lepton \pt and the angular distance between the lepton and the nearest jet.
The lepton identification efficiency is parametrized as a function of lepton $\pt$ and $\eta$.
Using these parametrizations, the efficiencies measured in $\Z$ events can be applied to the kinematically different \W{}+jets and \ttbar events.
The remaining differences in the $\pt$ and $\eta$ spectra of the signal and control regions are found to be smaller than $10\%$ in the simulation.

Leptons can be out of the acceptance because either their transverse momentum is too small or they are emitted in the forward direction.  
Electrons and muons from $\tau$ decays in particular tend to have low momentum, while the additional neutrinos add to the \MHT of the event.
The ratio $R_{\mbox{\tiny Accept}}$ of events with out-of-acceptance leptons to  those within the acceptance is estimated using simulation.
The same muon control sample described above is used, weighted by $R_{\mbox{\tiny Accept}}$ and corrected for the isolation and identification efficiencies, to estimate the background from out-of-acceptance leptons.

The dominant uncertainties on the lost-lepton prediction arise from the statistical uncertainties of both the control sample and the $\Z$ sample from which the lepton efficiencies are measured.
A systematic uncertainty is assigned to the kinematic differences between the control and signal regions that remain after the lepton-efficiency parametrization.
The residual presence of QCD, $\Z$, or diboson events in the control sample is taken into account as a systematic uncertainty. Finally, the systematic uncertainty due to the use of the simulation in the acceptance correction is considered.
All uncertainties are summarized in Table~\ref{tab:systematics}. The total systematic uncertainty amounts to approximately $18\%$.

\begin{table}[htb]
\begin{center}
\caption{Systematic uncertainties for the prediction of the lost-lepton background from the $\mu$+jets control sample.}
\label{tab:systematics}
\begin{tabular}{|l|rr|}
\hline \rule{0pt}{12pt}%
 Isolation \& identification eff.                    & $-13\%$ & $+14\%$  \\
 Kinematic differences between $\W$, $\ttbar$, $\Z$ samples  & $-10\%$ & $+10\%$  \\
 SM background in $\mu$ control sample               & $-3\%$  & $+0\%$   \\
 MC use for acceptance calculation                   & $-5\%$  & $+5\%$   \\
\hline \rule{0pt}{12pt}%
 Total  systematic uncertainty                       & $-17\%$ & $+18\%$  \\
\hline
\end{tabular}
\end{center}
\end{table}

The prediction from this method applied to the muon control sample collected using the same \HT triggers as for the search is compared in Table~\ref{tab:LostLeptonResult} to a prediction from simulated \W{}+jets and \ttbar events using the same method, and to the direct prediction from two different MC simulations.
When applied to simulation, the method reproduces within the uncertainties the direct expectations from the simulation.
Using the prediction from data after the baseline selection, about $50\%$ more events are predicted than expected from the \PYTHIA and \MADGRAPH simulated samples. The difference is due to the generator parameter tune
in the MC samples that were used to perform the comparison. %and due to the normalization of the \W component to the LO cross section.

\begin{table}[htb]
\begin{center}
\caption{Estimates of the number of lost-lepton background events from data and simulation for the baseline and search selections, with their statistical and systematic uncertainties.}
\label{tab:LostLeptonResult}
\begin{tabular}{|l|r@{\,}l|r@{\,}l|r@{\,}l|}
  \hline \rule{0pt}{12pt}%
  & \multicolumn{2}{c|}{Baseline}  & \multicolumn{2}{c|}{High-\MHT} & \multicolumn{2}{c|}{High-\HT}  \\
  & \multicolumn{2}{c|}{selection} & \multicolumn{2}{c|}{selection} & \multicolumn{2}{c|}{selection} \\
  \hline \rule{0pt}{12pt}%
  Estimate from data           & $33.0 $&$ \pm \, 5.5 \, {}^{+6.0}_{-5.7}$  & $4.8 $&$ \pm \, 1.8 \, {}^{+0.8}_{-0.6}$ & $10.9 $&$ \pm \, 3.0 \, {}^{+1.7}_{-1.7}$ \\
  Estimate from MC (\PYTHIA)   & $22.9 $&$ \pm \, 1.3 \, {}^{+2.7}_{-2.6}$  & $3.2 $&$ \pm \, 0.4 \, {}^{+0.5}_{-0.5}$ & $7.2  $&$ \pm \, 0.7 \, {}^{+1.1}_{-1.1}$ \\
  MC expectation (\PYTHIA)     & $23.6 $&$ \pm \, 1.0$                      & $3.6 $&$ \pm \, 0.3$                     & $7.8  $&$ \pm \, 0.5$ \\
  Estimate from MC (\MADGRAPH) & $22.9 $&$ \pm \, 1.4 \, {}^{+2.9}_{-2.8}$  & $2.7 $&$ \pm \, 0.4 \, {}^{+0.4}_{-0.4}$ & $5.4  $&$ \pm \, 0.5 \, {}^{+0.7}_{-0.6}$ \\
  MC expectation (\MADGRAPH)   & $23.7 $&$ \pm \, 0.8$                      & $3.4 $&$ \pm \, 0.3$                     & $6.5  $&$ \pm \, 0.5$ \\
\hline
\end{tabular}
\end{center}
\end{table}

\subsection{The \texorpdfstring{$\W/\ttbar\to\tau_{\rm h}\textrm{+X}$}{W/tt->tau\_h+X} background estimation}
\label{sec:wtop_hadronictau}

Hadronically decaying tau leptons constitute an important second component of the \W and \ttbar background. In this section a method is described to estimate the hadronic-$\tau$ background from a $\mu$+jets control sample, mainly composed of $\wmunubr$+jets and $\ttbarMuNu$ events.
This muon control sample is selected from data collected with single-muon triggers, ensuring independence from the hadronic activity in the event.
Events are required to have exactly one muon with $\pt>20\GeVc$ and $|\eta|<2.1$ and to satisfy the identification and isolation requirements described in Section~\ref{sec:eventselection}.

Jets from tau leptons are characterized by a low multiplicity of particles, typically a few pions and neutrinos.
The hadronic properties of events in the hadronic-$\tau$ background are identical to those of the muon control sample, except for the fraction of the $\tau$-jet energy deposited in the calorimeters.
To account for this difference, each muon in the control sample is replaced by a $\tau$ jet.
The momentum of this $\tau$ jet is obtained by scaling the muon momentum by a factor obtained from a simulated energy response distribution that models the fraction of visible momentum as a function of the true lepton momentum~\cite{Ztautau,Wtautau}. The extra jet is then taken into account when applying the selection cuts to obtain the hadronic-$\tau$ background prediction from these modified events. In order to probe the full response distribution, this procedure is repeated multiple times for each event.
 
A correction is applied for the kinematic and geometric acceptances of the muons in the control sample.
It is determined
by applying a muon smearing procedure to events in \W and \ttbar simulated samples with a muon from \W decay passing the muon kinematic selection, and comparing the resulting yield to the one obtained using all muons from \W decay in the same events.
The resulting correction factor is $0.84\pm0.05$ for the baseline and high-\mht selection, and $0.89\pm0.05$ for the high-\HT selection.
A second correction takes into account the muon trigger, reconstruction, and isolation efficiencies. The same procedure described in Section~\ref{sec:wtop_lostlepton} is followed.
A correction is also applied for the relative branching fractions of \W decays into muons or hadronic $\tau$ jets. For the simulated events a factor of 0.65 is used in the generation of the events and as the correction factor, while for data a factor of 0.69 is applied~\cite{PDG}.

The procedure for predicting the hadronic-$\tau$ background was tested on simulated \W and \ttbar events and reproduces the direct results from the simulation of genuine hadronic tau leptons from \W and \ttbar{} decays within uncertainties. For the baseline selection this uncertainty amounts to 12\% and 3\% for the \W and \ttbar samples, respectively.
The evaluation of the statistical uncertainty on the prediction needs special attention owing to the multiple sampling of the response distribution. This uncertainty is evaluated with a set of pseudo-experiments using the so-called bootstrap technique~\cite{tEFR82a}.

The systematic uncertainties and their impact on the prediction are summarized in
Table~\ref{tab:hadrtauuncert}. The possible difference between data and simulation for the $\tau$ energy distribution is taken into account as a systematic uncertainty, estimated by scaling the visible energy fraction by 
3\%~\cite{PFT-11-001}, resulting in a variation in the \MHT prediction of 2\%.
Possible SM background contamination in the muon control region comes from \zmumu,
$\ttbar / \W {\rm +X} \to \tau \nu  {\rm +X} \to \mu \nu {\rm +X}$, and from QCD multijet events.  The first two are subtracted using the MC simulation,
while the QCD multijet background is studied using an orthogonal control sample of events with non-isolated muons. The main source of background is $\W \rightarrow \tau\nu \to \mu\nu$, estimated to be 10\% of the total control sample.
The number of $\W / \ttbar \rightarrow\tau_{\rm h} {\rm +X}$ events predicted in data using this method is summarized 
in Table~\ref{tab:TAUpred} for the different signal regions.

\begin{table}[htb]
  \begin{center}
    \caption{Systematic uncertainties for the hadronic-$\tau$ background prediction from the $\mu$+jets control sample for the baseline and search selections.}
    \label{tab:hadrtauuncert}
    \begin{tabular}{|l|c|c|c|}
      \hline \rule{0pt}{12pt}%
       & Baseline  & High-\MHT & High-\HT \\
       & selection & selection & selection \\
      \hline \rule{0pt}{12pt}%
      $\tau$ response distribution & $2\%$     & $2\%$     & $2\%$     \\
      Acceptance                & $+6\%$/$-5\%$ & $+6\%$/$-5\%$ & $+6\%$/$-5\%$ \\
      Muon efficiency in data   & $1\%$     & $1\%$     & $1\%$     \\
      SM backgr. subtraction    & $5\%$     & $5\%$     & $5\%$     \\
      \hline 
    \end{tabular}
  \end{center}
\end{table}

\begin{table}[htb]
  \begin{center}
    \caption{Predicted number of hadronic-$\tau$ background events from data and simulation for the baseline and search selections, with their statistical and systematic uncertainties.}
    \label{tab:TAUpred}
    \begin{tabular}{|l|l|l|l|}
      \hline \rule{0pt}{12pt}%
      & \multicolumn{1}{c|}{Baseline}  & \multicolumn{1}{c|}{High-\MHT} & \multicolumn{1}{c|}{High-\HT}  \\
      & \multicolumn{1}{c|}{selection} & \multicolumn{1}{c|}{selection} & \multicolumn{1}{c|}{selection} \\
      \hline \rule{0pt}{12pt}%
      $\W/\ttbar\rightarrow\tau_{\rm h}$ estimate from data &
        $22.3 \pm 4.0 \pm 2.2$ &
        $ 6.7 \pm 2.1 \pm 0.5$ &
        $ 8.5 \pm 2.5 \pm 0.7$ \\
      $\W/\ttbar\rightarrow\tau_{\rm h}$ MC expectation &
        $19.9 \pm 0.9$ &
        $ 3.0 \pm 0.4$ &
        $ 5.5 \pm 0.5$ \\
      \hline
    \end{tabular}
  \end{center}
\end{table}

\section{QCD background estimation}
\label{sec:qcd}

Two
methods are employed to estimate the QCD multijet contamination in this
analysis.
The ``re\-ba\-lance-and-smear'' (\RplusS) method estimates the multijet background
directly from the data. This
method predicts the full kinematics in multijet events, while being unaffected by events with true missing transverse momentum, including the potential presence of a signal. Crucial inputs to this method are the jet energy resolutions, which are measured from data, including the non-Gaussian tails.
The ``factorization method'' provides an alternative prediction for the QCD
multijet background, based on the extrapolation from a lower-\MHT control region to the high-\MHT search region using the correlation between \MHT and an angular variable.

\subsection{The rebalance-and-smear method}
\label{sec:qcd_rebalancesmear}

Large missing transverse momentum arises in QCD multijet events when one or more jets in the event have a jet energy response far from unity, where the jet energy response is the ratio of the transverse momentum of the reconstructed jet over the one which would result from measuring perfectly the four-momenta of the particles in the jet (``particle jet'').
The \RplusS method is essentially a simplified simulation where the jet energy response is modelled by a parametrized resolution function, which is used to smear a sample of ``seed events'' obtained from data and consisting of ``seed jets'' that are good estimators of the true particle-jet momenta.

The
seed events are produced in the ``rebalance'' step with an inclusive multijet data sample as input.
Using the resolution probability distribution $r$, these seed events are constructed by adjusting the jet momenta in events with $n$ jets given the likelihood $\rsL=\prod_{i=1}^{n}r(\ppt i{\reco}|\ppt i{\true})$, where $\ppt i{\reco}$ and $\ppt i{\true}$ are the reconstructed and true jet transverse momentum, respectively. The likelihood is maximized as a function of $\ppt i{\true}$, subject
to the transverse momentum balance constraint $\sum_{i=1}^{n}\pptV i{\true}+\trueSoft=0$.
Here, all clustered objects with $\pt>10\GeVc$ are classified as jets and $\trueSoft$, which is the true momentum of the rest of the event, can be approximated by the measured
$\recoSoft$ that comprises all particles not included in the jets.
In other words, in the rebalancing step all of the jet momenta are adjusted, given the measured jet momenta and the jet resolution functions, to bring the event into transverse momentum balance.
This forces events with genuine high \mht from neutrinos or other undetected particles to be similar to well-balanced QCD-like events. As such, \ttbar, \W{}+jets, and \Z{}+jets events, and also contributions from new physics, if any, have negligible impact on the background prediction since their production rate is much smaller than the QCD multijet production rate.

Most of the events in the seed sample consist of jets with responses well within the core of the resolution distribution. Because of this, the Gaussian resolution model is sufficient in the computation of the likelihood. A correction is needed, though, because jets near masked ECAL cells have an energy response below unity, as discussed in Section~\ref{sec:qcd_jetres}, and hence get systematically rebalanced to too-low energies.
To mitigate the resulting bias, each event is randomized in $\phi$ after rebalancing, such that well-rebalanced events dominate everywhere. A second correction to the rebalancing procedure is applied to account for the migration of reconstructed events towards higher $\HT$ due to residual resolution smearing effects. An empirical term is included in the likelihood function to compensate for this migration.
This correction induces less than a 5\% change in the resulting distributions from data.

\begin{figure}[htb]
\centering
\includegraphics[width=0.49\textwidth]{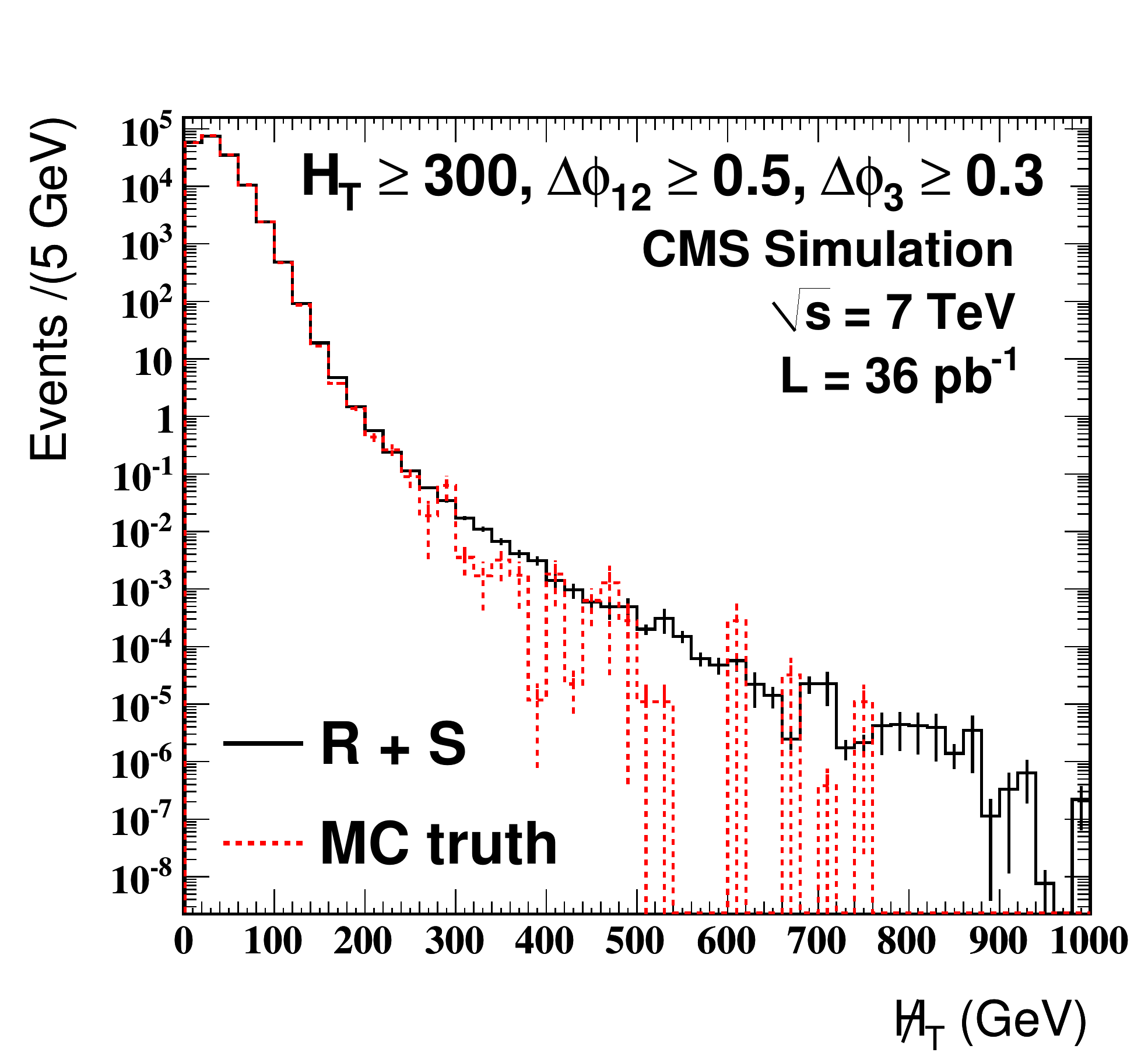}
\includegraphics[width=0.49\textwidth]{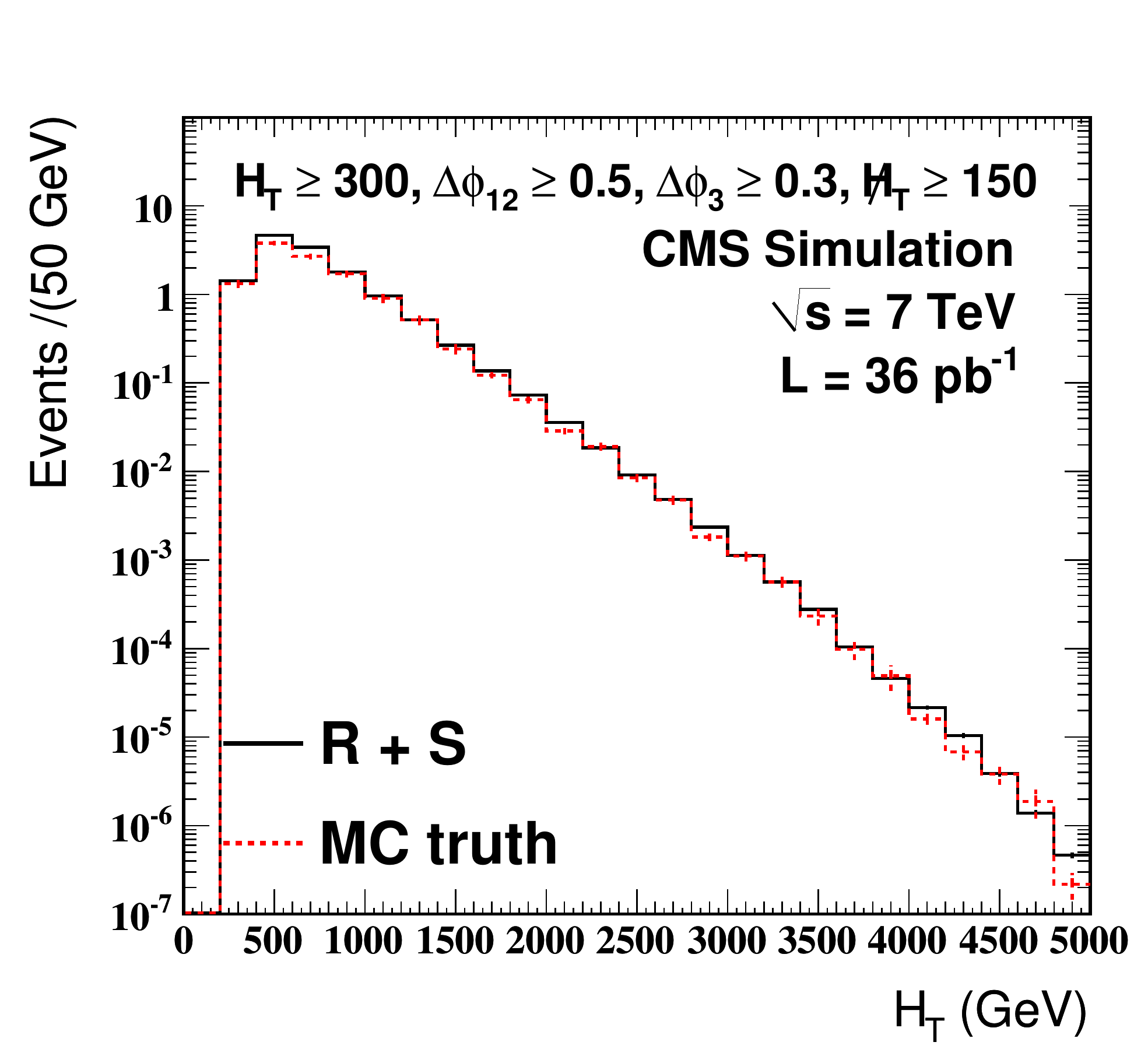} 
\caption{\label{fig:rebalance-smear-performance}The (left) \mht and (right) \HT distributions from the \RplusS method applied to simulation events, compared to the actual MC distribution (MC truth), for events passing $\geq3$~jets, $\HT\geq300\GeVc$, and $\DphiMHTjet{3}$ selections, and additionally $\MHT{} > 150 \GeVc$ for the right plot.}
\end{figure}

Next, the momentum of each seed jet is smeared using the jet resolution distribution.
The search requirements can be applied to the resulting smeared events to predict all event-by-event jet kinematic properties and correlations.
This allows for flexibility in the set of observables
used to define the search region, and in characterizing
an observed signal.
The distributions predicted by the \RplusS procedure are compared with those from MC simulation in Fig.~\ref{fig:rebalance-smear-performance}.
The predicted $\MHT$ and $\HT$ distributions are within 40\% of the actual MC distributions in the search regions; the corresponding numbers of events are listed in Table~\ref{tab:rebalance-smear-closure}.

\subsection{Jet response distributions}
\label{sec:qcd_jetres}

For smearing, and therefore the prediction of the \MHT spectrum, the full resolution functions including the non-Gaussian tails are used. The tails of the jet response function are of particular importance for the prediction of the QCD multijet background at high \MHT.

\begin{figure}[htb]
 \begin{center}
  \includegraphics[width=0.49\textwidth]{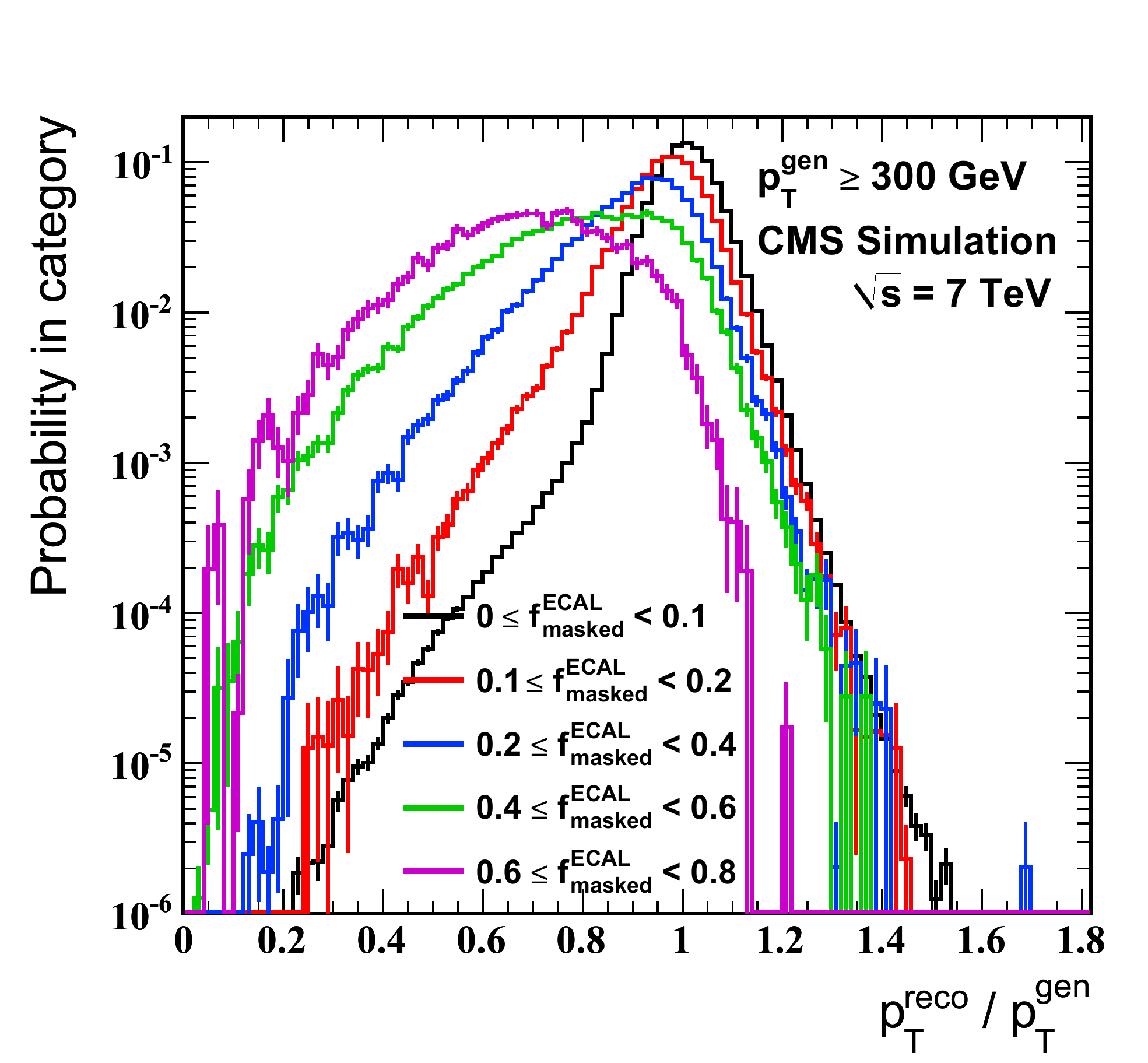}
  \hspace{\stretch{1}}
  \includegraphics[width=0.49\textwidth]{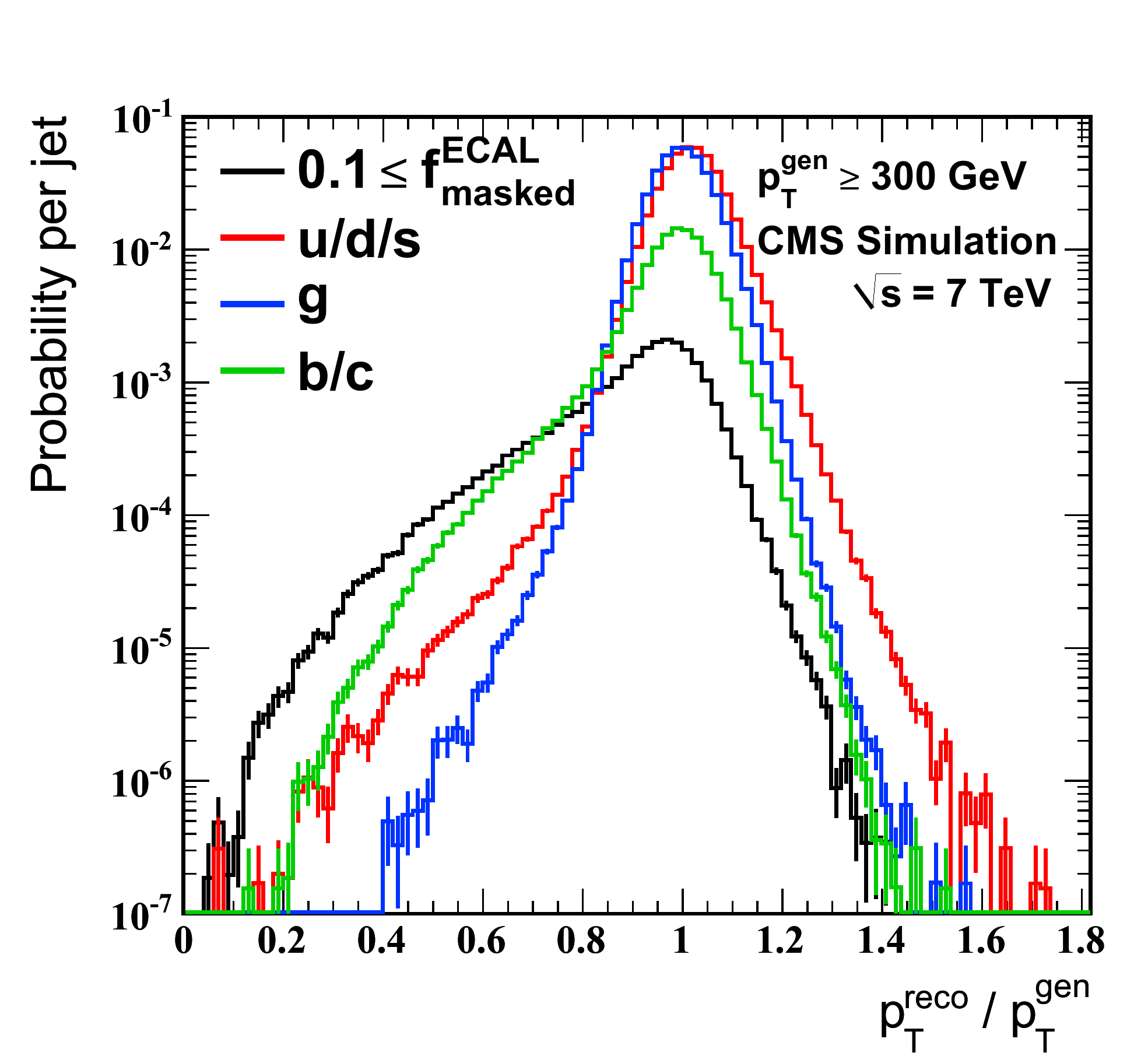}
  \caption{Ratio of the reconstructed jet transverse momentum to the generated transverse momentum for jets with $\pt^{\rm gen}\geq300\GeVc$. Distributions are shown for (left) different values of $\badFrac$ and (right) gluons and different quark flavours.}
  \label{fig:ptRes}
 \end{center}
\end{figure}

The jet momentum resolution functions are parametrized using simulated \PYTHIA dijet samples and adjusted to match
the measurements from data, as described below.
The response distributions are parametrized with respect to \pt and $\eta$.
Furthermore, an exceptionally low response arises at the specific $\eta-\phi$ locations where ECAL channels have been masked. This effect is taken into account by parametrizing the jet response as a function of the fraction $\badFrac$ of jet momentum lost in the masked area of the detector, computed using a template for the $\pt$-weighted distribution of particles as a function of the distance in $\eta$ and $\phi$ to the jet axis. The dependence of the jet resolution on $\badFrac$ is shown in Fig.~\ref{fig:ptRes} (left).
Finally, heavy-flavour ${\rm b}$ or ${\rm c}$ quarks and also gluons exhibit different jet resolution shapes than light jets, as shown in Fig.~\ref{fig:ptRes} (right).
At high jet $\pt$, decays of heavy-flavour hadrons into neutrinos become one of the dominant sources of significant jet energy loss. 
The jet resolution functions are determined for bottom, charm, gluon, and other light-flavour quarks separately.
The flavour dependence is then accounted for by using these resolution functions in the smearing procedure according to the flavour fractions from simulation. 

Two methods are used to measure from data a scaling factor for the Gaussian core of the jet momentum resolutions determined from simulation~\cite{jetrespas}.
At low \pt, $\gamma$+jet events are used because the photons are reconstructed with excellent energy resolution and the \pt balance makes the photons good estimators of the true \pt scale of the event.
At larger \pt, dijet events are used.
An unbinned maximum likelihood fit is performed on the dijet asymmetry, $(\pt^{\text{jet1}}-\pt^{\text{jet2}})/(\pt^{\text{jet1}}+\pt^{\text{jet2}})$, with random ordering of the two highest-\pt jets.
For both measurements the presence of additional jets in the event destroys the momentum balance and an extrapolation to no-additional-jet activity is performed.
These methods measure the core of the Gaussian resolution as a function of jet $\eta$ to be $5-10\%$ larger in data compared to simulation, with systematic uncertainties of similar size as the deviation. No significant dependence on the \pt of the jet is observed.

No significant non-Gaussian tails are observed in $\gamma$+jet events.
At higher \pt, the dijet asymmetry distributions show compatibility within uncertainties between the resolution tails from data and simulation.
Using the ratio of these asymmetry distributions in data and simulation, correction factors to the jet resolution tails from simulation are derived.

Both a scaling
of the response below (``low'' tail) and above unity (``high'' tail) can induce the same change in the
asymmetry distribution; the latter arising for instance from mismeasured track momenta in particle-flow jets. Therefore, the nominal resolution functions are obtained by equally scaling both the lower and upper tails of the resolution in order
to induce the observed scaling of the asymmetry tail. The envelope of the variations induced by only low- or high-tail scaling is taken as the systematic uncertainty band for the jet resolution distribution.

\subsection{Results of the rebalance-and-smear method}
\label{sec:qcd_rsresults}

The performance of the \RplusS procedure was validated using simulated \PYTHIA{} QCD multijet samples, without pileup interactions, where the parametrized response functions are derived from the same samples.
The predicted and expected number of events are summarized for several selections in Table~\ref{tab:rebalance-smear-closure}.
Before the $\MHT$ requirement, the prediction of the $\HT$ spectrum, the jet kinematics, and the jet-jet angular and $\pt$ correlation distributions agree within 10\% with the direct simulation. The $\MHT$ distribution shows a bias up to 40\%, which is mostly due to a dependency of the jet resolution on the presence of nearby jets. This is only of importance in the region of very high \MHT, however, where the QCD multijet contribution is negligible
compared to other backgrounds.

\begin{table}[htb]
\begin{center}
\caption{\label{tab:rebalance-smear-closure}Number of events passing the various event selections from the \PYTHIA{} multijet sample, the \RplusS method applied to the same simulated sample, and their ratio.
The uncertainties quoted are statistical only.}
\begin{tabular}{|l|r@{$\;\pm\;$}lr@{$\;\pm\;$}lr@{$\;\pm\;$}lr@{$\;\pm\;$}l|}
\hline \rule{0pt}{12pt}%
 & \multicolumn{2}{c}{Baseline selection} & \multicolumn{2}{c}{Baseline} & \multicolumn{2}{c}{high-\MHT} & \multicolumn{2}{c|}{high-\HT} \\
 & \multicolumn{2}{c}{No $\Dphi$ cuts} & \multicolumn{2}{c}{selection} & \multicolumn{2}{c}{selection} & \multicolumn{2}{c|}{selection} \\
\hline \rule{0pt}{12pt}%
N(\PYTHIA) & \hspace{5mm}$138.6$&$1.3$ & $11.4$&$0.4$ & $0.13$&$0.04$ & $8.46$&$0.32$\\
N(\RplusS) & $160.2$&$0.1$ & $13.2$&$0.1$ & $0.177$&$0.004$ & $9.57$&$0.04$\\
\hline \rule{0pt}{12pt}%
N(\RplusS)/N(\PYTHIA) & $1.16$&$0.01$ & $1.15$&$0.04$ & $1.4$&$0.4$ & $1.13$&$0.05$\\
\hline
\end{tabular}
\end{center}
\end{table}

The QCD multijet background is predicted using the inclusive data sample with events passing the same \HT triggers described in Section~\ref{sec:evseldata}. The \RplusS steps are then executed using the jet energy resolution functions and the core and tail scale factors described in Section~\ref{sec:qcd_jetres}. The background predictions are obtained by applying the event selection requirements to the \RplusS events. The rejection efficiency of events with large energy loss in masked ECAL channels is modelled using a parametrized per-jet probability from simulation.

In Table~\ref{tab:rebalance-smear-prediction} the number of predicted
events is listed for the baseline and search regions, along with the relevant systematic uncertainties. 
Corrections are applied to the background estimates for several known biases in the method, as summarized in Table~\ref{tab:rebalance-smear-prediction}.
The largest one pertains to the smearing step, and arises from ambiguities in how the jet resolution is defined and from limitations in the parametrization. It is obtained in simulation by comparing the prediction from smeared particle jets with the corresponding one from the detector simulation. The size of the difference is taken as both a bias correction and a systematic uncertainty.

\begin{table}
\begin{centering}
\caption{\label{tab:rebalance-smear-prediction} Number of QCD multijet events predicted with the \RplusS method, before and after bias corrections, along with all considered uncertainties and the type of uncertainty (uniform ``box''-like, symmetric or asymmetric Gaussian distribution). Effects in italics are the biases corrected for as described in the text, with
the full size of the bias taken as the systematic uncertainty.}
\begin{tabular}{|lc|ccc|}
\cline{1-5}
\multicolumn{1}{|c}{} &  & Baseline  & high-\MHT & high-\HT \\ \multicolumn{1}{|c}{} &  & selection & selection & selection \tabularnewline
\hline \rule{0pt}{12pt}%
\textbf{Nominal prediction} (events) & & 39.4 & 0.18 & 19.0\tabularnewline
\hline \hline
\emph{Particle jet smearing closure} & (box) & $+14\%$ & $+30\%$ & $\;\,+7\%$\tabularnewline
\hline 
\emph{Rebalancing bias} & (box) & $+10\%$ & $+10\%$ & $+10\%$\tabularnewline
\hline 
\emph{Soft component estimator} & (box) & $\;\,+3\%$ & $+19\%$ & $\;\,+4\%$\tabularnewline
\hline 
Resolution core & (asymmetric) & $\begin{array}{c}
+14\%\\
-25\%\end{array}$ & $\begin{array}{c}
+0\%\\
-52\%\end{array}$ & $\begin{array}{c}
+15\%\\
-21\%\end{array}$\tabularnewline
\hline 
Resolution tail & (asymmetric) & $\begin{array}{c}
+43\%\\
-33\%\end{array}$ & $\begin{array}{c}
+56\%\\
-78\%\end{array}$ & $\begin{array}{c}
+48\%\\
-34\%\end{array}$\tabularnewline
\hline 
Flavour trend & (symmetric) & $\;\,\pm1\%$ & $\pm12\%$ & $\pm0.3\%$\tabularnewline
\hline 
Pileup effects & (box) & $\;\,\pm2\%$ & $\pm10\%$ & $\;\,\pm2\%$\tabularnewline
\hline 
Control sample trigger & (box) & $\;\,-5\%$ & $\;\,-5\%$ & $\;\,-5\%$\tabularnewline
\hline 
Search trigger & (symmetric) & $\;\,\pm1\%$ & $\;\,\pm1\%$ & $\quad\, 0\%$\tabularnewline
\hline 
Lepton veto & (box) & $\;\,\pm5\%$ & $\pm0.05\%$ & $\pm0.2\%$\tabularnewline
\hline 
Seed sample statistics & (symmetric) & $\pm2.3\%$ & $\pm23\%$ & $\pm3.3\%$\tabularnewline
\hline 
\multicolumn{2}{|l|}{\rule{0pt}{12pt}Total uncertainty} & \;\;\;$51\%$ & $\;\;\;64\%$ & $\;\;\;49\%$\tabularnewline
\hline \hline \rule{0pt}{12pt}%
\textbf{Bias-corrected prediction} (events) & & $29.7 \pm 15.2$ & $0.16 \pm 0.10$ & $16.0 \pm 7.9$ \tabularnewline
\hline
\end{tabular}
\par\end{centering}
\end{table}

A second bias is intrinsic to the rebalancing procedure, and is studied by iterating the \RplusS method. A first iteration $\rsN{}^{1}$ of the method gives a sample of pure QCD multijet events with known true jet resolution, i.e., by construction the one used in the smearing step. Performing a second iteration $\rsN{}^{2}$ of the method on this $\rsN{}^{1}$ sample, using the same resolutions, provides a closure test of just the rebalancing part when compared to the input $\rsN{}^{1}$ events. The degree of non-closure is measured to be 10\%, which is also assigned as a systematic uncertainty.

The same $\rsN{}^{2} / \rsN{}^{1}$ procedure is employed to study the bias
caused by using $\recoSoft$ as an estimator of $\trueSoft$. The true value of $\trueSoft$ in the second iteration is equal to the \MHT value calculated from the rebalanced jets in the first iteration.
The difference between the \mbox{$\rsN{}^{2}$} predictions with
$\recoSoft$ and $\trueSoft$ as input is used as a third bias correction, with corresponding systematic uncertainty.

The largest systematic effect arises from uncertainties on the jet momentum resolution.
The measurement uncertainties on the core resolutions and non-Gaussian tails, discussed in Section~\ref{sec:qcd_jetres}, are propagated by repeating the \RplusS prediction with resolution inputs varied within these uncertainties. 
Another systematic uncertainty comes from the flavour-dependent para\-me\-tri\-za\-tion of the jet resolutions. It is evaluated as the difference between the use of \PYTHIA and \MADGRAPH simulated samples to derive the ${\rm b}$- and ${\rm c}$-quark content parametrization. These MC generators have heavy-flavour fractions that differ by roughly 25\% for bottom and 50\% for charm quarks.
Nevertheless, the difference in the resulting background prediction is very small in the high-\HT search regions, and the QCD multijet contribution is negligible in the high-\MHT search region.

The effect of pileup is studied by performing the \RplusS prediction with a subset of events with exactly one reconstructed primary vertex. The relative difference between this prediction and the one obtained from the inclusive sample is taken as a systematic uncertainty.

Other smaller uncertainties arise from the event selection. 
A potential loss of events due to the \HT trigger requirement on the events that enter the rebalancing is quantified by comparing the prediction made with the small number of events collected with a low-\pt single-jet trigger. 
A conservative upper bound of 5\% on this uncertainty is taken. Another uncertainty arises from the need to predict the number of smeared events that pass the search trigger. The \HT triggers used were measured on data to be fully efficient with respect to events passing the offline cuts, and the statistical upper bound from this measurement is taken as a systematic uncertainty for the low-\HT selections.
Finally, the lepton veto has an uncertainty that is estimated as the full size of the rejection rate for QCD multijet events in a \PYTHIA event sample with pileup conditions representative of the data. The large size of this uncertainty for the baseline search region 
is due to a near-100\% statistical uncertainty induced by an MC sample with a very small equivalent luminosity.

Variations within one standard deviation or within upper and lower bounds are performed for each systematic effect, and the corresponding differences in the predictions are quoted in Table~\ref{tab:rebalance-smear-prediction}. Estimated shapes of the probability distribution of each uncertainty are also listed; uncertainties
that are estimated as upper bounds on possible effects are assumed to have a uniform ``box''-like distribution.
The statistical uncertainty is associated with the size of the seed event sample. As prescribed by the bootstrap method~\cite{tEFR82a}, an ensemble of pseudo-datasets is selected randomly from the original seed sample, allowing repetition. The ensemble spread of predictions made from these pseudo-datasets is taken as the statistical uncertainty.

After correcting for biases, the \RplusS prediction and systematic uncertainties are combined via the procedure explained in Section~\ref{sec:resultslimits}, which takes properly into account
non-Gaussian distributed uncertainties. The mean and r.m.s. deviation of the resulting distributions of the expected number of multijet
background events for the baseline and search selections are taken as the central values and uncertainties of the final \RplusS prediction, as given in the last row of Table~\ref{tab:rebalance-smear-prediction}. These central values are slightly shifted compared to the nominal bias-corrected values owing to the asymmetrically distributed uncertainties.

\subsection{The factorization method}
\label{sec:qcd_abcd}

Because of the importance of estimating the QCD multijet background, an independent approach is used as a cross-check. The factorization method uses the observables \MHT and \DeltaPhi, of which the latter is the minimum azimuthal angle between the \MHT direction and the three leading jets, to predict the number of events in the signal region of high \MHT and large \DeltaPhi{} from the sideband regions where one or both variables are small. As \MHT and \DeltaPhi{} are not independent observables, their correlation is measured in the low-\MHT region by means of the ratio $r(\MHT)$ of the number of events with large $\DeltaPhi$ to the number with small $\DeltaPhi$. The number of background events is estimated from the extrapolation of $r$ to the high-\MHT signal region.

The parametrization of $r(\MHT)$ is chosen empirically, with two different ones being used. 
The first parametrization, the Gaussian model, predicts a Gaussian distribution for \DeltaPhi{}, assuming all jets, except the most mismeasured jet, to have an energy response following a Gaussian resolution function.
The width of this distribution as a function of \MHT is described both in simulation and data by a falling exponential function, from which the functional form for \mbox{$r(\MHT)$} is derived.
An additional constant term, determined from a \MADGRAPH QCD multijet simulation, is added to $r(\MHT)$ to keep it more-nearly constant at high values of \MHT. A large value of $\HT$ is further required to suppress events with low-\pt jets at low \MHT.
This method results in a prediction for a lower limit on the number of expected QCD multijet background events in the signal region, since any non-Gaussian tails in the \DeltaPhi{} resolutions result in a larger estimate.

As an alternative to the Gaussian resolution model, $r(\MHT)$ is parametrized as an exponential plus the same constant term used in the Gaussian model. The same \HT cut is applied. The extrapolation to high \mbox{\MHT} leads to a larger $r(\MHT)$ value than observed in the simulation.
Various systematic variations of simulated QCD samples show that the true yield of the QCD multijet background is between the predictions from the two parametrizations.

The dominant uncertainty on the prediction is the statistical uncertainty from the data in the control region and the statistical uncertainties on the fit parameters.
A systematic uncertainty arises from the constant term at high \MHT for both models, which is estimated to be $+11\%$/$-6\%$ from a variety of different simulated samples.
Further systematic uncertainties come from the SM background contamination in the control regions, $+4\%$/$-8\%$, and from the $\HT$ requirement discussed above, $+0\%$/$-11\%$.

The predictions for the QCD multijet background from the two parametrizations are given in Table~\ref{tab:FactorizationResult} for the three different selections.
The final background estimate is taken as the average of the two model predictions, with half the difference assigned as an additional systematic uncertainty and added linearly to the uncertainty on the combination.
The results are in agreement with the predictions using the \RplusS method.

\begin{table}[htb]
\begin{center}
\caption{Predictions for the number of QCD multijet background events using the factorization method with two different parametrizations and their combination, for the baseline and search selections, with their statistical and systematic uncertainties.}
\label{tab:FactorizationResult}
\begin{tabular}{|l|l|l|l|}
  \hline \rule{0pt}{12pt}%
  Method & \multicolumn{1}{c|}{Baseline selection} & \multicolumn{1}{c|}{High-\MHT} & \multicolumn{1}{c|}{High-\HT}  \\
         & \multicolumn{1}{c|}{Baseline selection} & \multicolumn{1}{c|}{selection} & \multicolumn{1}{c|}{selection} \\
  \hline \rule{0pt}{12pt}%
  Gaussian model    & $19.0 \pm 1.6 \, {}^{+7.2}_{-6.5}$   & $0.3 \pm 0.1 \, {}^{+0.4}_{-0.2}$ & $13.0 \pm 1.3 \, {}^{+4.9}_{-4.4}$  \\
  Exponential model & $31.4 \pm 2.4 \, {}^{+7.0}_{-6.9}$   & $0.5 \pm 0.1 \, {}^{+0.2}_{-0.2}$ & $21.6 \pm 2.0 \, {}^{+4.8}_{-4.8}$  \\
  Combined          & $25.2 \pm 2.4 \, {}^{+13.2}_{-13.1}$ & $0.4 \pm 0.1 \, {}^{+0.3}_{-0.3}$ & $17.3 \pm 2.0 \, {}^{+9.1}_{-9.2}$  \\
  \hline
\end{tabular}
\end{center}
\end{table}

\section{Results and interpretation}
\label{sec:sensitivity}

\subsection{Results and limits} \label{sec:resultslimits}

The number of events observed in data and the event yields predicted by the different background estimation methods
are summarized in Table~\ref{tab:FinalEventYields} for the three
different selections. The total background is calculated summing the QCD \RplusS, the \znunubr{}+jets from photons, and the \W/\ttbar lost-lepton and hadronic-$\tau$ estimates. No excess of events is observed in either the
\mbox{high-\MHT} or high-\HT search regions.

\begin{table}[htb]
\begin{center}
\caption{Predicted number of background events from the different estimates for the baseline and search
selections, their total, and the corresponding number of events observed in data. The background combination is performed as explained in the text.
The uncertainties shown include both statistical and systematic uncertainties. The last line gives the 95\% confidence level (CL) upper limit on the number of possible signal events.}
\label{tab:FinalEventYields}
{
\begin{tabular}{|l|r@{}l|r@{}l|r@{}l|}
\hline \rule{0pt}{12pt}%
  Background process & \multicolumn{2}{c|}{Baseline} & \multicolumn{2}{c|}{High-\MHT} & \multicolumn{2}{c|}{High-\HT}  \\
         & \multicolumn{2}{c|}{selection} & \multicolumn{2}{c|}{selection} & \multicolumn{2}{c|}{selection} \\
\hline \rule{0pt}{12pt}%
  $\znunubr$+jets ($\gamma$+jets method)     & $ 26.3 \;$ & $\pm \;  4.8$ & $ 7.1  \;$ & $\pm \; 2.2 $ & $ 8.4 \;$ & $\pm \; 2.3$ \\
  $\W/\ttbar\to \e,\mu$+X                    & $ 33.0 \;$ & $\pm \;  8.1$ & $ 4.8  \;$ & $\pm \; 1.9 $ & $10.9 \;$ & $\pm \; 3.4$ \\
  $\W/\ttbar\to \tau_{\mbox{\tiny h}}$+X  & $ 22.3 \;$ & $\pm \;  4.6$ & $ 6.7  \;$ & $\pm \; 2.1 $ & $ 8.5 \;$ & $\pm \; 2.5$ \\
  QCD multijet (\RplusS method)              & $ 29.7 \;$ & $\pm \; 15.2$ & $ 0.16 \;$ & $\pm \; 0.10$ & $16.0 \;$ & $\pm \; 7.9$ \\
  \hline \rule{0pt}{12pt}%
  Total background                           & $111.3 \;$ & $\pm \; 18.5$ & $18.8  \;$ & $\pm \; 3.5 $ & $43.8 \;$ & $\pm \; 9.2$ \\
  \hline \rule{0pt}{12pt}%
  Observed in data                           & $111\;$\hspace{3mm} &      & $15\;$\hspace{3mm} &       & $40\;$\hspace{3mm} &     \\
  \hline \rule{0pt}{12pt}%
  95\% CL upper limit on signal              & $ 40.4 \;$ &               & $ 9.6  \;$ &               & $19.6 \;$ &              \\
\hline
\end{tabular}
}
\end{center}
\end{table}

In order to derive limits on new physics, the expected number of signal events for the event selections are
estimated using simulated signal samples, taking into account uncertainties on the
event selection, theoretical uncertainties related to the event generation, and an overall luminosity
uncertainty.
Many of these uncertainties have a dependence on the event kinematics,
and hence are model dependent.

The largest experimental contribution to the uncertainties arises from the model-dependent jet energy scale and resolution uncertainties. These amount to 8\% for the LM1 benchmark point.
Smaller uncertainties are due to the lepton veto and the trigger. For the former a 2\% uncertainty is determined for LM1; for the latter a
conservative uncertainty of $1\%$ is assigned. The inefficiency of the rejection of events with energy in masked ECAL cells is
determined to be about $1.5\%$ for the LM1 benchmark point. This full inefficiency is taken as the uncertainty, even though the ECAL masked-channel simulation reproduces well the effect in
data~\cite{METJINST}. For other event cleaning procedures, possible inefficiencies are determined in a \mbox{low-\MHT}
data control region to be negligible. Also the possible effect from the presence of additional
pileup interactions corresponding to the LHC 2010 data-taking conditions was investigated and found
to be insignificant. On the theoretical side, all uncertainties considered are model-dependent.
The largest one is associated with the factorization
and renormalization scale uncertainties on the next-to-leading-order cross-section corrections, yielding a 16\% uncertainty for the LM1 point.
Smaller contributions come from uncertainties on the parton distribution functions and initial-state
radiation, respectively 3\% and 2\% for LM1. Final-state radiation uncertainties are found to be negligible. Finally, a luminosity uncertainty of 4\% is accounted for~\cite{lumi-moriond},
along with the statistical uncertainty on the simulated signal samples, which is about 2\% for the LM1 sample.

The probability distributions corresponding to each
uncertainty source, whether Gaussian, bifurcated Gaussian, Poisson, or box shaped,
are convolved using a numerical integration MC technique to
obtain the probability 
distributions for each background and for the overall background estimation.
The presence of several sources of uncertainties makes the overall combination
quite Gaussian 
in shape, as expected from the central limit theorem. The resulting
distribution is fitted to a Gaussian function, and the mean and standard deviation are
used as the central value and uncertainty in the limit calculations described in the
following sections. This last step is applied in order to obtain the best
symmetric approximation to a distribution with a residual asymmetry.

\subsection{Interpretation within the CMSSM} \label{sec:limitsmsugra} 

\begin{figure}[htb]
\includegraphics[width=0.49\textwidth]{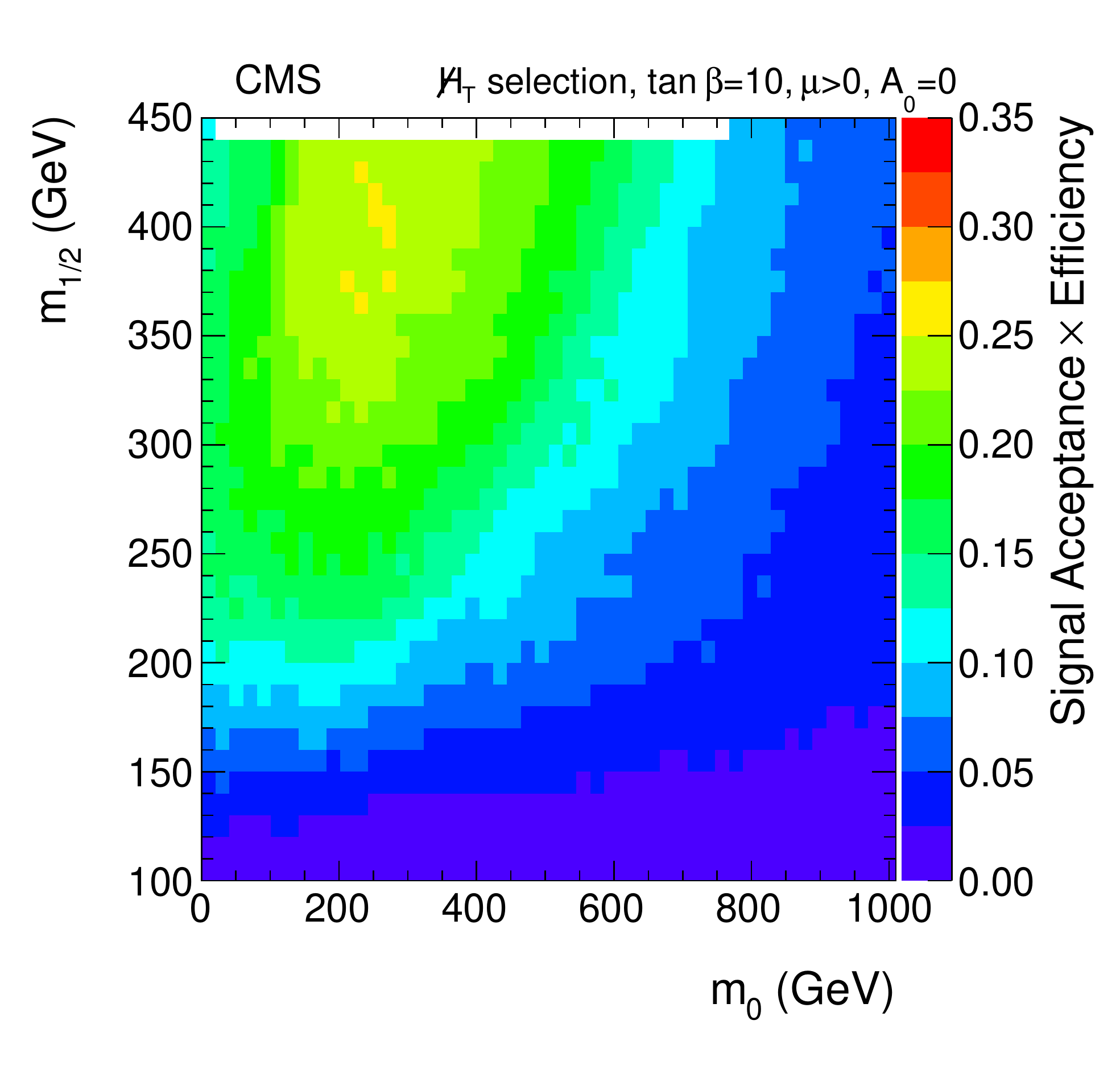}
\hspace{\stretch{1}}
\includegraphics[width=0.49\textwidth]{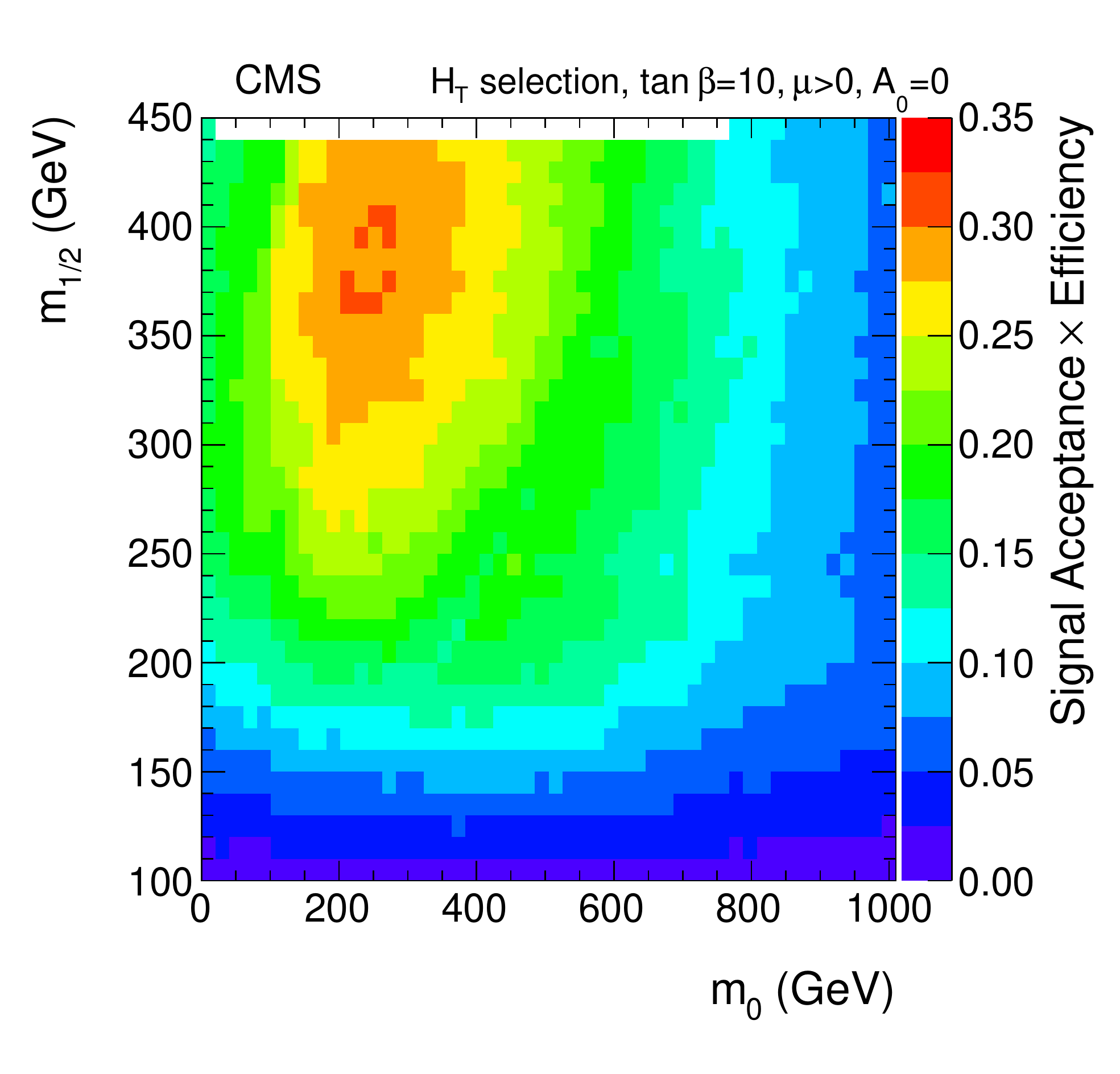}
\caption{Total signal efficiency for the \MHT (left) and \HT (right) selections, as a function of $m_0$ and $m_{1/2}$. The other CMSSM
parameters are  $\tan\beta=10$, $\mu>0$, and $A_0=0$.}
\label{fig:msugraeffs}
\end{figure}

The parameters $m_0$ and $m_{1/2}$ of the CMSSM
are varied in $10 \GeV$ steps for
three different values of $\tan\beta=3$, $10$, and $50$. Leading-order {\sc
IsaJet}~\cite{Isajet} signal cross sections are used and corrected by next-to-leading-order
$K$ factors calculated using {\sc Prospino}~\cite{Prospino}.  The total signal efficiency, including geometrical acceptance and selection efficiency,
varies over the CMSSM phase space, being in the range $20-30\%$ for
the high-\HT selection and $10-20\%$ for the high-\MHT selection, as shown in Fig.~\ref{fig:msugraeffs}.

The expected upper limits on the CMSSM cross section are calculated using the background estimate from data
under the no-signal hypothesis.
For the determination of the observed upper limit the signal contamination in the background estimate is corrected for.
In the isolated-muon control region of the lost-lepton and hadronic-$\tau$ methods, the signal contamination is calculated and removed from the background estimate for each CMSSM parameter point. For both
selections, the signal contributions to the background estimate are $2-3$ events for the lost leptons and $1-2$ events for
the hadronic tau decays.  The signal contamination in the $\gamma$+jets control region is found to be negligible. The
QCD multijet background estimation with the \RplusS method is not affected by signal contamination.

\begin{figure}[htb]
\hspace{-3mm}
\includegraphics[angle=0,width=0.52\textwidth]{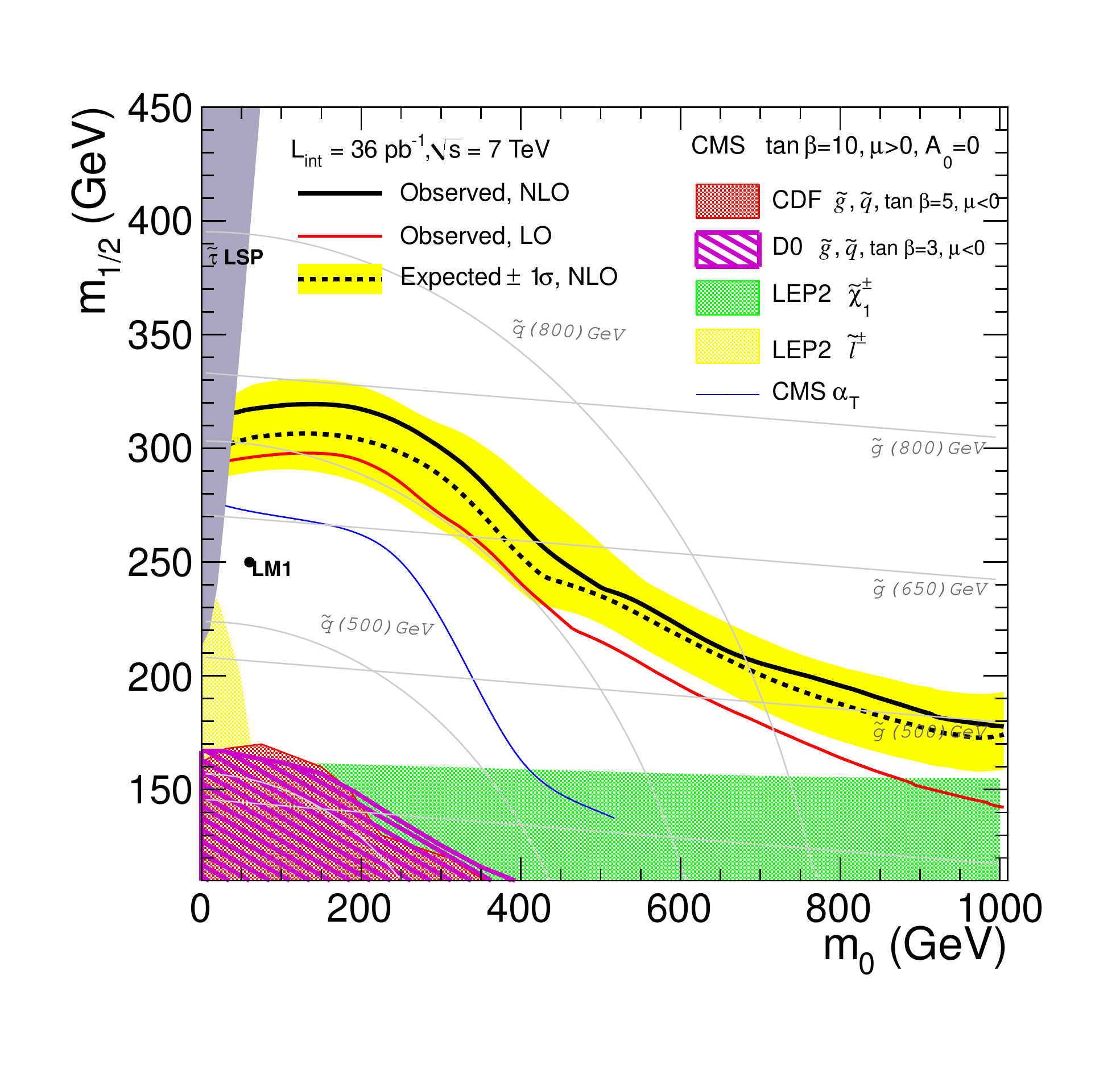}
\hspace{-3mm}
\includegraphics[angle=0,width=0.5\textwidth]{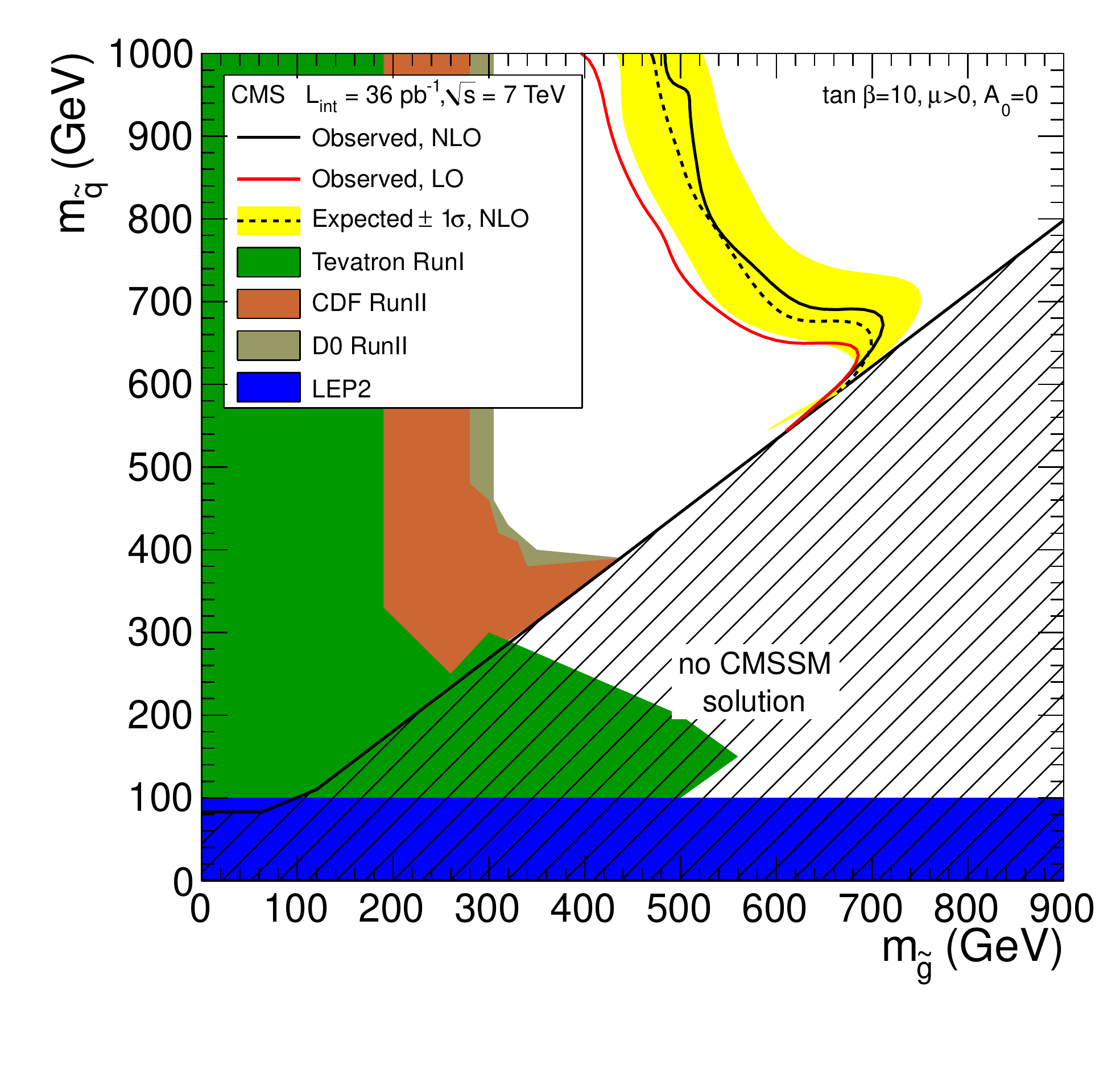}
\caption{The expected and observed $95\%$ CL upper limits in the CMSSM $m_0$--$m_{1/2}$ (left) and gluino--squark (right) mass planes for LO and NLO cross sections.  The $\pm1$ standard deviation ($\sigma$) band corresponds
to the expected limit.  The contours are the combination of the \HT and the \MHT selections
such that the contours are the envelope with respect to the best sensitivity. The CMSSM
parameters are  $\tan\beta=10$, $\mu>0$, and $A_0=0$. The limit from the earlier CMS analysis is shown as a blue line and limits from other experiments as the shaded regions. For the area labeled ``$\tilde{\tau}$ LSP'' the stau becomes the LSP. The LM1 SUSY benchmark scenario is shown as a point.}
\label{fig:msugralimits}
\end{figure}

The modified frequentist procedure CLs~\cite{aread,tjunk} with a likelihood ratio test-statistic is used for the limit calculation. 
In Fig.~\ref{fig:msugralimits} the observed and expected CLs 95\% confidence level (CL) upper limits 
are shown in the CMSSM $m_0$-$m_{1/2}$ (left) and the gluino-squark (right) mass planes for $\tan\beta=10$,
$\mu>0$, and $A_0=0$. 
The contours are the envelope with respect to the best sensitivity
of both the \HT and the \MHT search selections. For $m_0\lesssim 450 \GeV$ the \mbox{\MHT} selection is more powerful, while for
large $m_0$ the \HT selection is more important. A previously published search by CMS for supersymmetry in hadronic events~\cite{RA1} using
the event shape observable $\alpha_T$~\cite{Randall:2008rw} is shown for
reference.
The $\alpha_T$ analysis aims at the best possible
removal of the QCD multijet background, and is particularly powerful for small jet multiplicities and high missing transverse
energy.  Because of the high signal selection efficiency in a large fraction of the phase space, and in spite of the 
larger background compared to the $\alpha_T$
selection, the analysis presented here is able to improve the limits previously set by the $\alpha_T$ analysis.

\subsection{Interpretation with Simplified Model Spectra} \label{sec:simplifiedmodels}

Models for new physics can also be studied in a more generic manner using a simplified model spectra (SMS) approach~\cite{Alwall:2008ag,Alwall:2008zz,SMS}.
Simplified models are designed to characterize experimental data in terms of a
small number of basic parameters. 
They exploit the fact that at the LHC the final-state kinematics of
events involving strongly produced massive new particles are largely determined
by the parton distribution functions and phase-space factors associated with two- and three-body decays.
Using these simplified models, the experimental results can then be translated into any desired framework.

\begin{figure}[htb]
  \begin{center}
      \includegraphics[width=0.3\textwidth,angle=90]{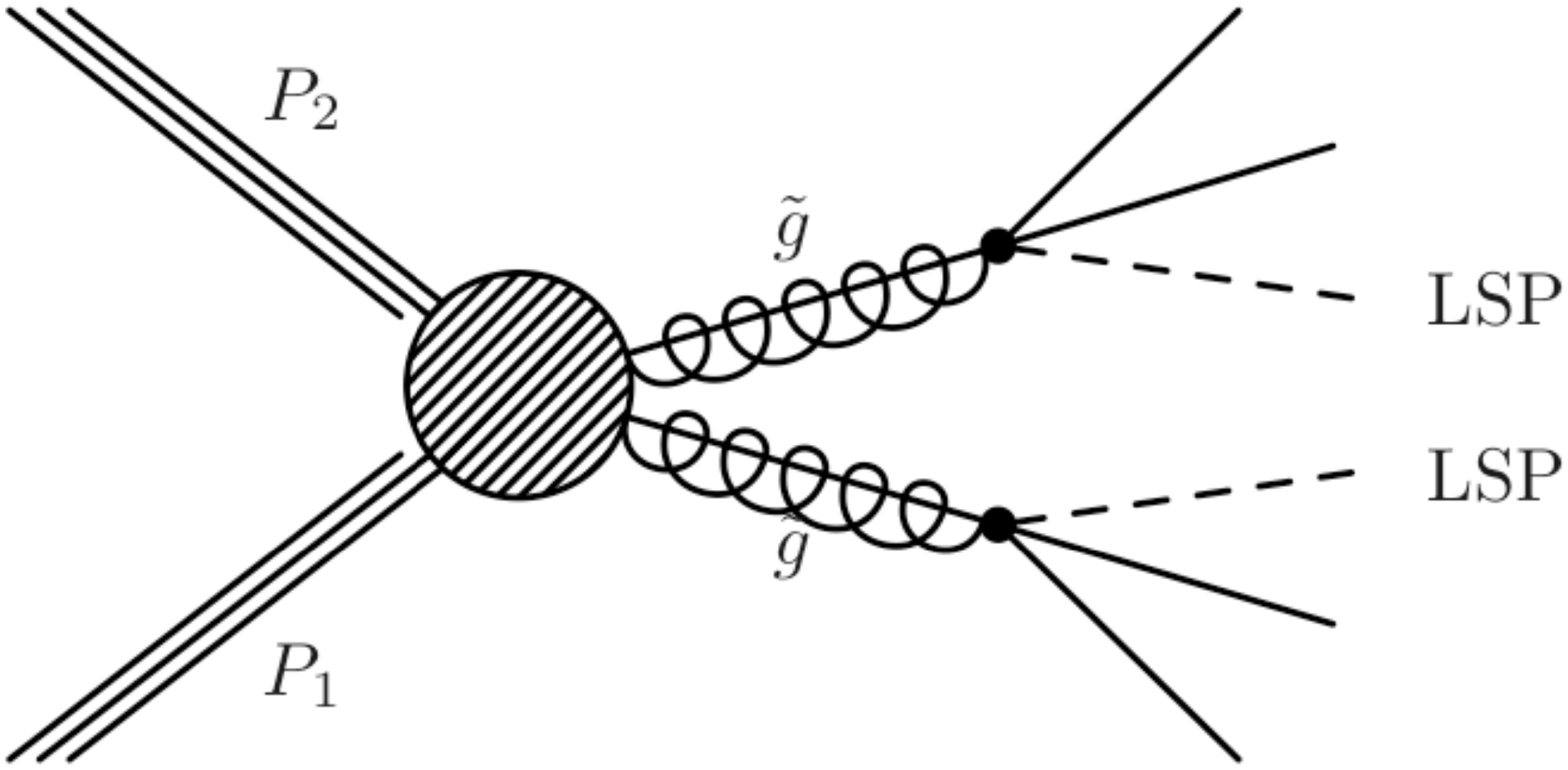}
      \includegraphics[width=0.3\textwidth,angle=90]{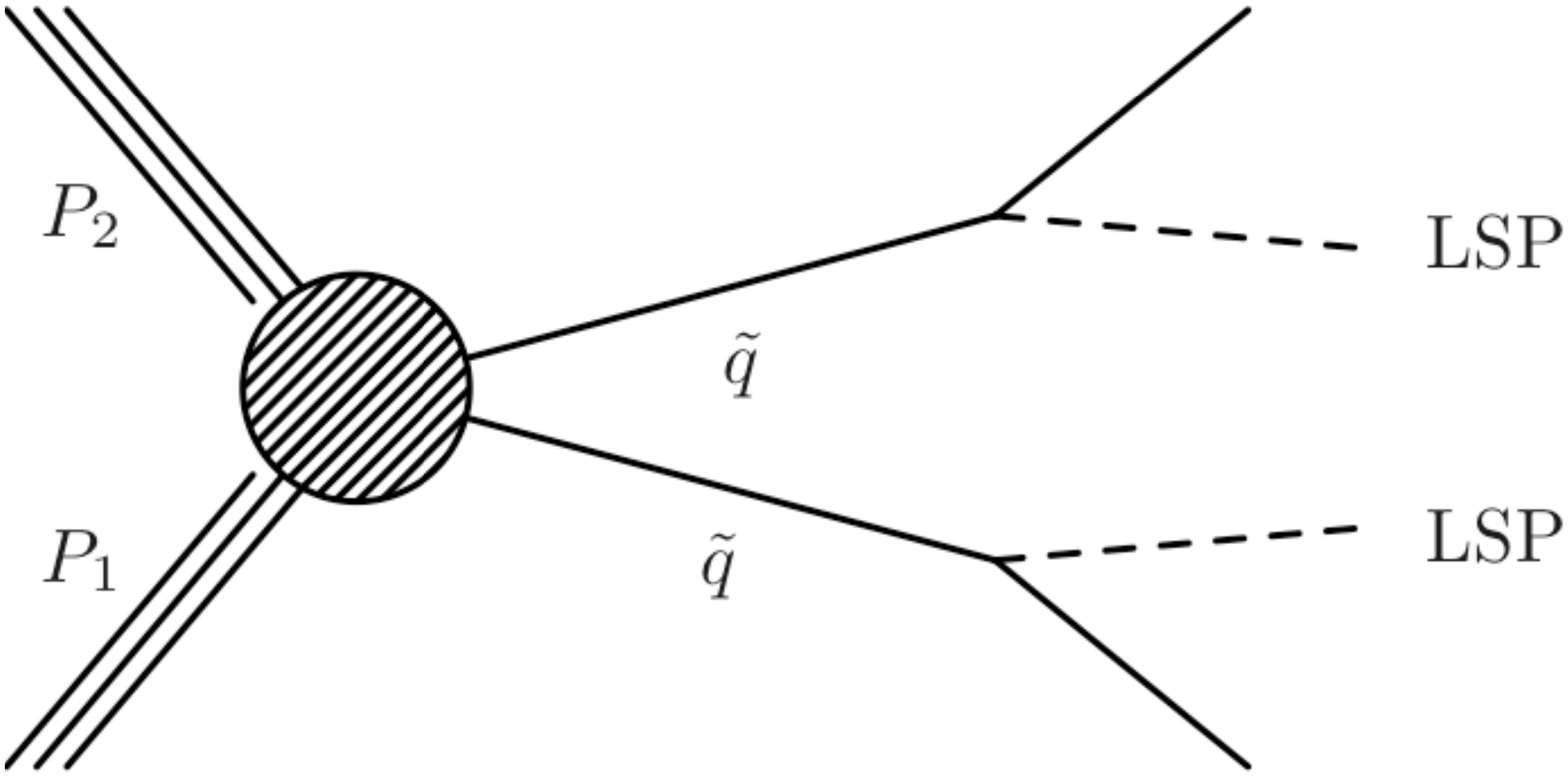}
      \vspace{-5mm}
	\caption{Diagrams of the studied simplified models. Left: gluino pair production; right: squark pair production.} %; bottom row: one single-step cascade decay for gluino pair production and squark pair production.}
    \label{fig:diagram}
  \end{center}
\end{figure}

For the simplified models used in this paper, it is assumed that the new particles
are strongly produced in pairs whose decay chains
ultimately result in a stable weakly interacting massive particle,
denoted as LSP.
The particles produced in the hard interaction can be identified as partners of
quarks and gluons. In SUSY these would be the squarks (\squark) and gluinos (\gluino).  Even though
the SMS are more generic, in the following everything is phrased for simplicity 
in terms of super-partner names.
Two benchmark simplified models are investigated for the number of jets and \MHT signature in this analysis:
pair-produced gluinos, where each gluino directly decays to two light quarks and the
LSP, and pair-produced squarks, where each squark decays to one light quark and the LSP.
In Fig.~\ref{fig:diagram} the respective diagrams for these simplified models are
drawn. To limit the set of SMS studied, only a few are chosen that can bracket
the kinematic properties of the different final states. For this reason the gluino-squark
associated production is neglected.

The simplified models are simulated with the \PYTHIA generator~\cite{pythia}, the CTEQ6L1 parton distribution functions~\cite{Pumplin:2002vw}, and the parametrized CMS detector simulation. For each topology, samples are generated for a range of masses of the particles
involved, and thus more mass splittings are explored than in the CMSSM, where
the ratio of the gluino and the LSP masses is approximately fixed.

In the following, the measured cross section upper limits are compared to a typical reference next-to-leading-order cross section from {\sc Prospino}~\cite{Prospino}. In the case of squark pair production this reference cross section corresponds to the squark-antisquark
cross section with four light flavours included, with the gluinos becoming nearly decoupled at $3 \TeV$. This cross section is used to
convert upper limits on the production cross section to reference limits on new-particle masses.

\begin{figure}[htb]
  \begin{center}
    \includegraphics[width=0.49\textwidth]{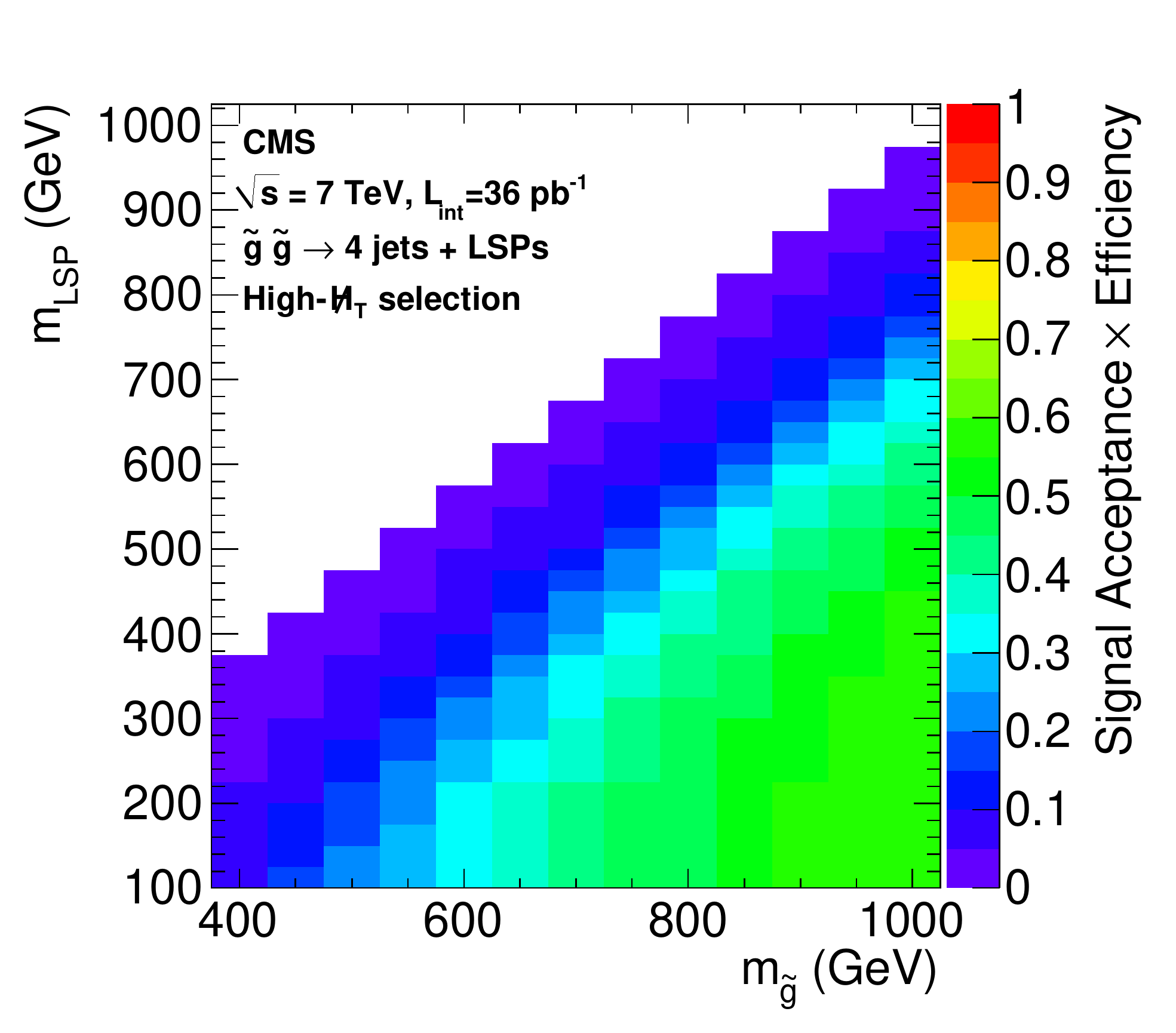}
    \includegraphics[width=0.49\textwidth]{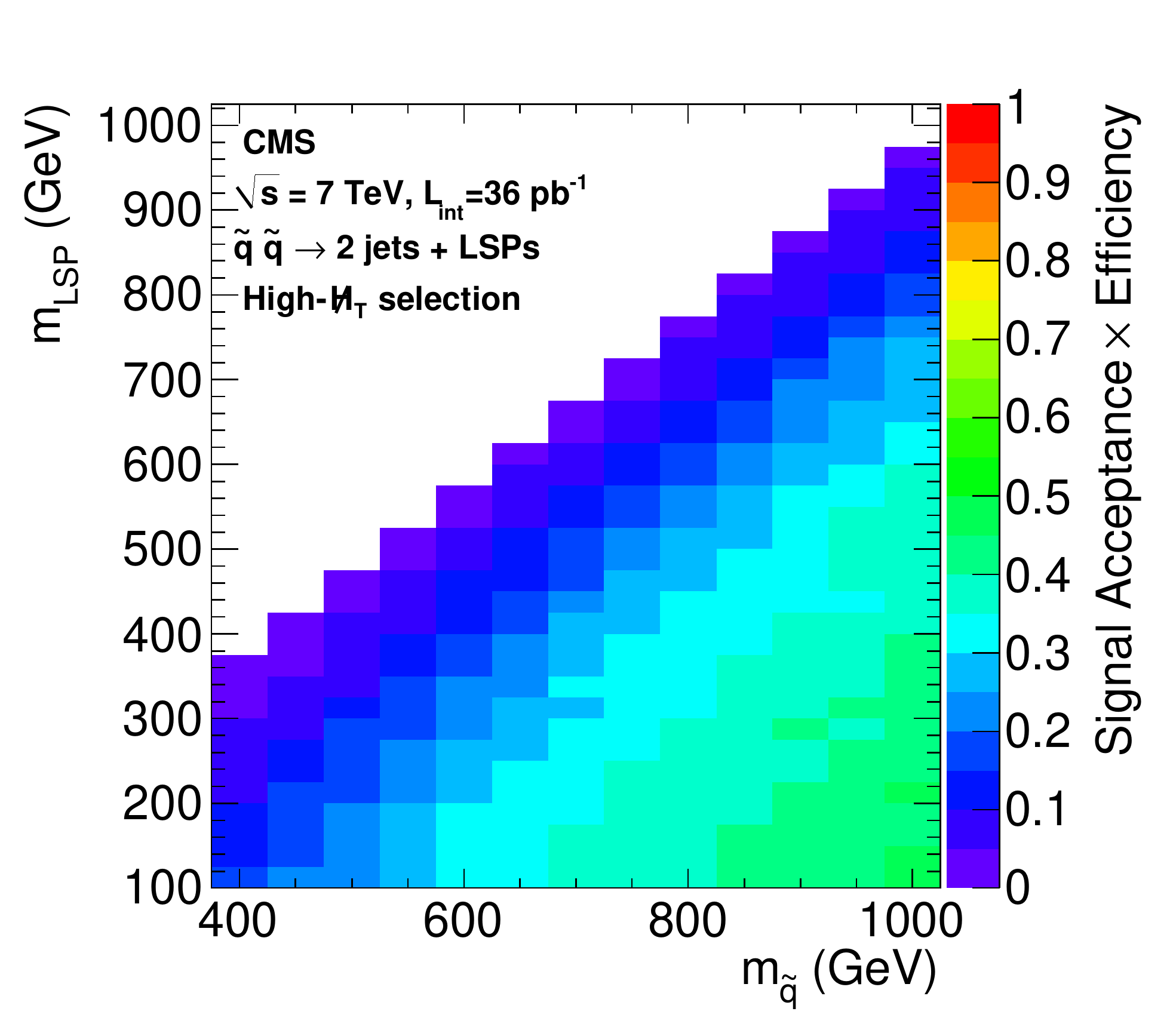}
    \caption{Total high-\MHT selection efficiency for gluino (left) and squark (right) production as a function of the gluino (left) or squark (right) mass and the LSP mass.}
    \label{fig:smstotseleff}
  \end{center}
\end{figure}

In Fig.~\ref{fig:smstotseleff} the total signal efficiency of the high-\MHT selection, including geometrical acceptance and selection efficiency, is shown within the simplified model space for gluino and squark pair production, as a function of the gluino (left) or squark mass (right) and the LSP mass. Only the lower half of the plane is filled because the model is only valid when the gluino or squark masses are larger than the mass of the LSP.
The signal selection efficiency increases for higher gluino and squark masses, and is low on the diagonal, where the mass splitting is small and jets are produced with lower transverse momentum.

The limit calculation in the SMS space is performed using a Bayesian framework with a flat prior for the signal~\cite{PDG}.
The same sources of uncertainties affecting the signal geometrical acceptance and selection efficiency are incorporated for each scan point as for the CMSSM interpretation, namely the jet energy scale and resolution, the lepton veto, the cleaning including the veto on large energy loss in masked ECAL cells, the trigger, the initial- and final-state radiation, the parton distribution functions, the luminosity, and the statistical uncertainty. 
The renormalization and factorization scale uncertainties do not apply here because they only influence the normalization of the reference cross section. The presence of signal events in the background sample is not considered, since the studied SMS processes do not produce prompt leptons or photons, and since the \RplusS method is insensitive to such contamination.

\begin{figure}[htb]
  \begin{center}
    \includegraphics[width=0.49\textwidth]{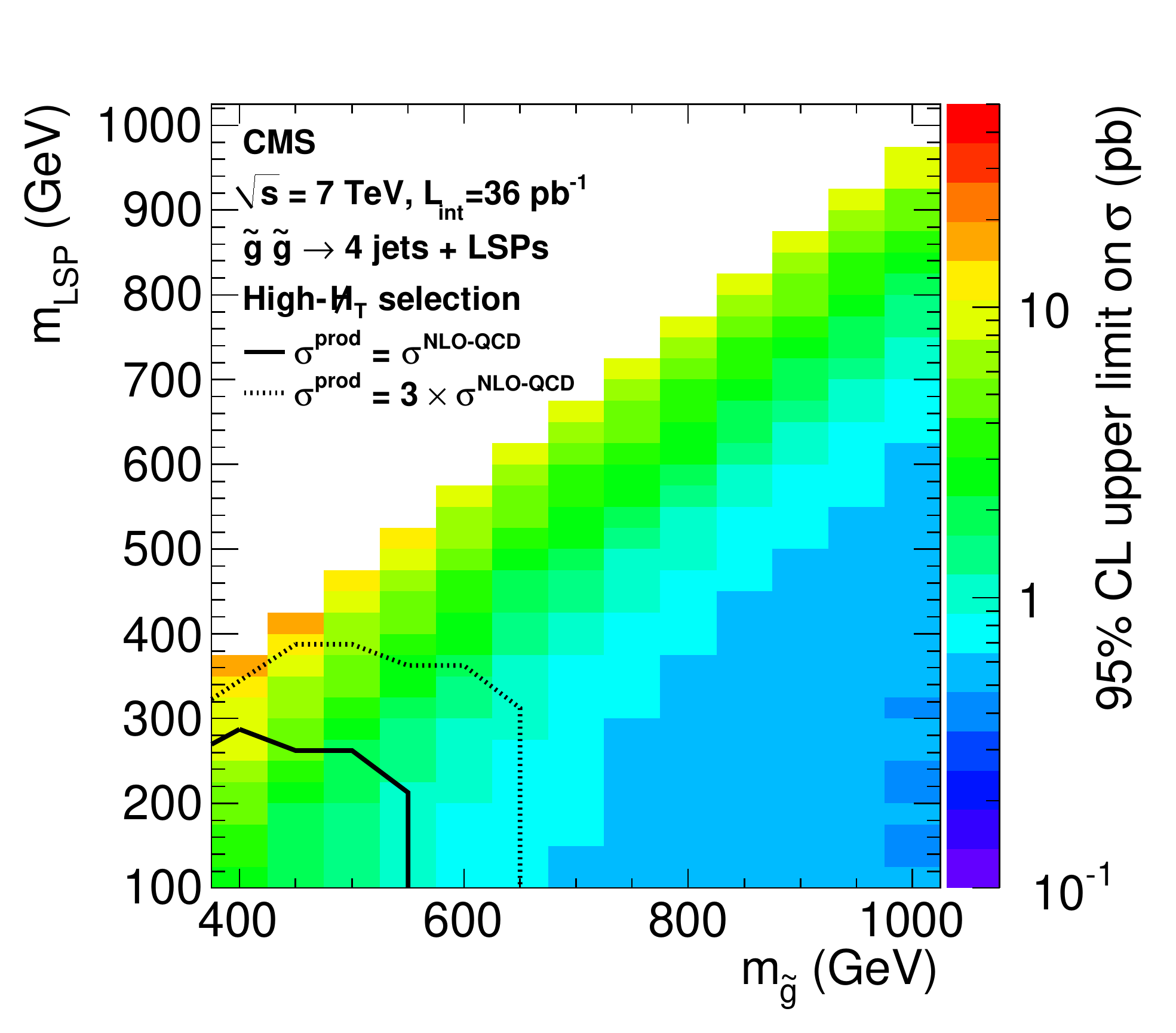}
    \includegraphics[width=0.49\textwidth]{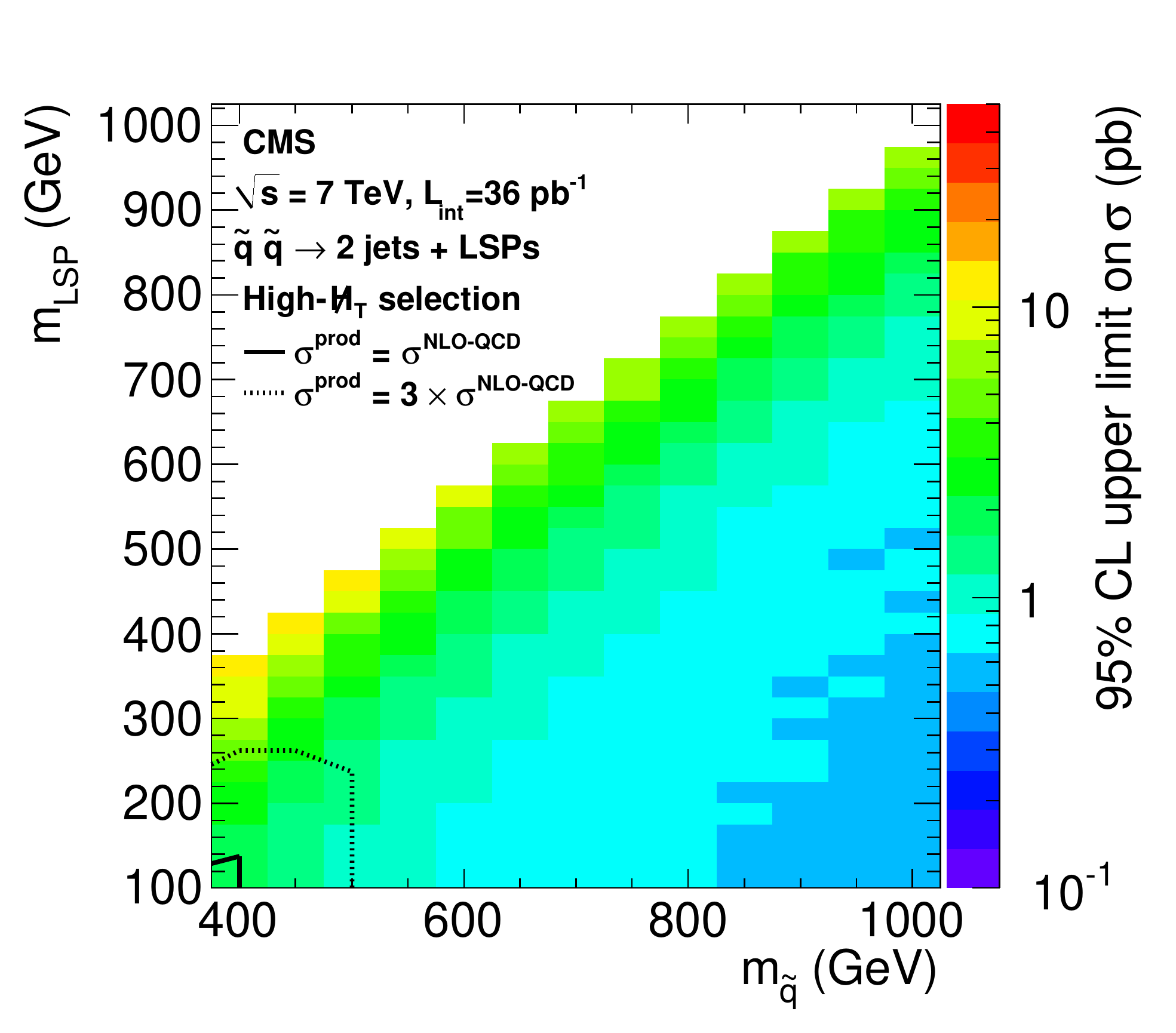}
    \caption{95\% CL upper limits on the gluino (left) and squark (right) pair-production cross sections for the high-\MHT selection, as a function of the gluino (left) or squark (right) mass and the LSP mass. The contours where the reference cross section and three times this cross section can be excluded are shown.}
    \label{fig:SMSlimit}
  \end{center}
\end{figure}

In Fig.~\ref{fig:SMSlimit} the exclusion 95\% CL upper limits on the production cross sections are presented for the \mbox{high-\MHT} search selection. This selection is found to be more sensitive than the high-\HT search selection for both considered simplified model spectra. 
Using this model-independent representation with the simplified model spectra, these upper limits on the cross section can be translated into a limit on any complete model such as SUSY.

\section{Conclusions}
\label{sec:conclusions}

An inclusive search for new physics has been presented using events with a multijet signature with large missing transverse momentum. The
observed event yield is consistent with the SM background contributions,
arising mainly from $\znunubr$+jets, $\wlnubr$+jets,
\ttbar{} including a \W that decays leptonically, and QCD multijet production. These SM contributions
were estimated directly from the data using several novel techniques, giving a minimal reliance on simulation. The overall uncertainty on the resulting total background prediction is dominated by the statistical uncertainty.

In the absence of an excess of events above the expectation, upper limits are derived in the CMSSM parameter space. In $R$-parity conserving CMSSM with $A_0=0$, $\mu>0$, and $\tan\beta=10$, a 95\% CL upper limit on the production cross section in the range between $2$ and $3\pb$ is obtained, depending on the squark and gluino masses considered. Gluino masses below $500 \GeVcc$ are excluded at 95\% CL for squark masses below $1\,000 \GeVcc$.
Similar results are obtained for other $\tan\beta$ values.
The results are also more generically interpreted in the context of simplified models where final
states are described by the pair production of new particles which decay either
to one or two jets and a dark-matter candidate escaping detection. We obtain a 95\% CL upper limit on the production cross section for
such new particles in the range between $0.5$ and $30 \pb$, depending on the masses of
the new particles in the decay chains.

\section*{Acknowledgements}
\label{sec:ack}

We thank the members of the {\sc BlackHat} collaboration for their input and fruitful discussions concerning vector boson production with jets at the LHC.

\hyphenation{Bundes-ministerium Forschungs-gemeinschaft Forschungs-zentren} We also wish to congratulate our colleagues in the CERN accelerator departments for the excellent performance of the LHC machine. We thank the technical and administrative staff at CERN and other CMS institutes. This work was supported by the Austrian Federal Ministry of Science and Research; the Belgium Fonds de la Recherche Scientifique, and Fonds voor Wetenschappelijk Onderzoek; the Brazilian Funding Agencies (CNPq, CAPES, FAPERJ, and FAPESP); the Bulgarian Ministry of Education and Science; CERN; the Chinese Academy of Sciences, Ministry of Science and Technology, and National Natural Science Foundation of China; the Colombian Funding Agency (COLCIENCIAS); the Croatian Ministry of Science, Education and Sport; the Research Promotion Foundation, Cyprus; the Estonian Academy of Sciences and NICPB; the Academy of Finland, Finnish Ministry of Education and Culture, and Helsinki Institute of Physics; the Institut National de Physique Nucl\'eaire et de Physique des Particules~/~CNRS, and Commissariat \`a l'\'Energie Atomique et aux \'Energies Alternatives~/~CEA, France; the Bundesministerium f\"ur Bildung und Forschung, Deutsche Forschungsgemeinschaft, and Helmholtz-Gemeinschaft Deutscher Forschungszentren, Germany; the General Secretariat for Research and Technology, Greece; the National Scientific Research Foundation, and National Office for Research and Technology, Hungary; the Department of Atomic Energy and the Department of Science and Technology, India; the Institute for Studies in Theoretical Physics and Mathematics, Iran; the Science Foundation, Ireland; the Istituto Nazionale di Fisica Nucleare, Italy; the Korean Ministry of Education, Science and Technology and the World Class University program of NRF, Korea; the Lithuanian Academy of Sciences; the Mexican Funding Agencies (CINVESTAV, CONACYT, SEP, and UASLP-FAI); the Ministry of Science and Innovation, New Zealand; the Pakistan Atomic Energy Commission; the State Commission for Scientific Research, Poland; the Funda\c{c}\~ao para a Ci\^encia e a Tecnologia, Portugal; JINR (Armenia, Belarus, Georgia, Ukraine, Uzbekistan); the Ministry of Science and Technologies of the Russian Federation, the Russian Ministry of Atomic Energy and the Russian Foundation for Basic Research; the Ministry of Science and Technological Development of Serbia; the Ministerio de Ciencia e Innovaci\'on, and Programa Consolider-Ingenio 2010, Spain; the Swiss Funding Agencies (ETH Board, ETH Zurich, PSI, SNF, UniZH, Canton Zurich, and SER); the National Science Council, Taipei; the Scientific and Technical Research Council of Turkey, and Turkish Atomic Energy Authority; the Science and Technology Facilities Council, UK; the US Department of Energy, and the US National Science Foundation.

Individuals have received support from the Marie-Curie programme and the European Research Council (European Union); the Leventis Foundation; the A. P. Sloan Foundation; the Alexander von Humboldt Foundation; the Associazione per lo Sviluppo Scientifico e Tecnologico del Piemonte (Italy); the Belgian Federal Science Policy Office; the Fonds pour la Formation \`a la Recherche dans l'Industrie et dans l'Agriculture (FRIA-Belgium); the Agentschap voor Innovatie door Wetenschap en Technologie (IWT-Belgium); and the Council of Science and Industrial Research, India.

\bibliography{auto_generated}   % will be created by the tdr script.

\cleardoublepage\appendix\section{The CMS Collaboration \label{app:collab}}\begin{sloppypar}\hyphenpenalty=5000\widowpenalty=500\clubpenalty=5000\textbf{Yerevan Physics Institute,  Yerevan,  Armenia}\\*[0pt]
S.~Chatrchyan, V.~Khachatryan, A.M.~Sirunyan, A.~Tumasyan
\vskip\cmsinstskip
\textbf{Institut f\"{u}r Hochenergiephysik der OeAW,  Wien,  Austria}\\*[0pt]
W.~Adam, T.~Bergauer, M.~Dragicevic, J.~Er\"{o}, C.~Fabjan, M.~Friedl, R.~Fr\"{u}hwirth, V.M.~Ghete, J.~Hammer\cmsAuthorMark{1}, S.~H\"{a}nsel, M.~Hoch, N.~H\"{o}rmann, J.~Hrubec, M.~Jeitler, W.~Kiesenhofer, M.~Krammer, D.~Liko, I.~Mikulec, M.~Pernicka, H.~Rohringer, R.~Sch\"{o}fbeck, J.~Strauss, A.~Taurok, F.~Teischinger, P.~Wagner, W.~Waltenberger, G.~Walzel, E.~Widl, C.-E.~Wulz
% ...
\vskip\cmsinstskip
% ...
1:~~Also at CERN, European Organization for Nuclear Research, Geneva, Switzerland\\
% ...

\end{sloppypar}
\end{document}